\documentclass[lettersize,journal]{IEEEtran}
\usepackage{amsmath}
\usepackage{times}
\usepackage{graphicx}
\usepackage{xcolor}
\usepackage[colorlinks=true, allcolors=blue]{hyperref}
\usepackage{booktabs}
\usepackage{array}
\usepackage{makecell}
\usepackage{pifont}
\usepackage{caption}
\usepackage{authblk}
\usepackage[T1]{fontenc}
\usepackage{multirow}
\usepackage{longtable}
\usepackage{makecell}

\usepackage{booktabs}
\usepackage{tabularx}
\usepackage{array}
\newcolumntype{Y}{>{\raggedright\arraybackslash}X}

\usepackage{stfloats}
\setcellgapes{4pt}
\makegapedcells

\usepackage[english]{babel}

\newtheorem{definition}{Definition}
\AtBeginDocument{%
  \fontdimen2\font=0.21em 
  \fontdimen3\font=0.1em  
  \fontdimen4\font=0.06em  
}

\begin{document}

\title{Agentic Services Computing}

\author{Shuiguang~Deng,~\IEEEmembership{Senior~Member,~IEEE,}
        Hailiang~Zhao,~\IEEEmembership{Member,~IEEE,}
        Ziqi~Wang,~\IEEEmembership{Student~Member,~IEEE,}
        Wenzhuo~Qian,
        Xiang~Ao,
        Guanjie~Cheng,
        Jianwei~Yin,
        Albert~Y.~Zomaya,~\IEEEmembership{Fellow,~IEEE,}
        Schahram~Dustdar,~\IEEEmembership{Fellow,~IEEE}
\thanks{Shuiguang Deng and Hailiang Zhao are corresponding authors.}
\thanks{Shuiguang Deng, Wenzhuo Qian, Xiang Ao, and Jianwei Yin are with College of Computer Science and Technology, Zhejiang University, Hangzhou 310027, China (e-mails: \{dengsg, qwz, aoxiangfly, zjuyjw\}@zju.edu.cn).}
\thanks{Hailiang Zhao, Ziqi Wang, and Guanjie Cheng are with School of Software Technology, Zhejiang University, Ningbo 315048, China (e-mails: \{hliangzhao, wangziqi0312, chengguanjie, zhenqin\}@zju.edu.cn).}
\thanks{Albert Y. Zomaya is with School of Computer Science, University of Sydney, Sydney, NSW 2006, Australia (e-mail: albert.zomaya@sydney.edu.au).}
\thanks{Schahram Dustdar is with the Distributed Systems Group, TU Wien, Vienna, Austria, and also with ICREA, Universitat Pompeu Fabra, Barcelona, Spain (e-mail: dustdar@dsg.tuwien.ac.at).}
}



\maketitle

\begin{abstract}
Services computing has evolved from Web services and microservices to cloud-native and serverless paradigms. These approaches established mature principles for describing, composing, deploying, operating, and governing reusable software functions. LLM-based agents now introduce a fundamentally different service form. Service value in this paradigm emerges not only from invoking predefined functions but also from delegating goals to autonomous entities. These entities understand context, use tools, collaborate with peers, and act across open environments. This shift raises a core question for services computing. How can goal-driven, stateful, tool-mediated, and accountable autonomous behavior be engineered and managed as a service? Recent studies on LLM agents and multi-agent systems provide important foundations. A clear service-centered research roadmap for this emerging paradigm nevertheless remains absent. This work introduces Agentic Services Computing (ASC) to address this gap. ASC extends services computing from managing reusable functional endpoints to engineering and governing autonomous service entities. It defines agentic services as service-oriented autonomous agents. Related research is organized through a lifecycle view that connects service objects, system structures, enabling infrastructure, evaluation metrics, application evidence, and open challenges. This service-centered perspective establishes a foundation for future service ecosystems. Autonomous agents can thus be systematically described, composed, delivered, monitored, audited, and evolved as first-class services.
\end{abstract}


\section{Introduction}\label{sec:intro}

Services computing has fundamentally evolved by continuously redefining what can be engineered, delivered, and managed as a service. From early service-oriented architectures that encapsulated software capabilities as interoperable units~\cite{papazoglou2003service,papazoglou2007service}, to microservices, cloud-native, and serverless paradigms that decomposed applications into independently deployable functions~\cite{newman2021building,baldini2017serverless,deng2024cloud}, the discipline has matured around a \textit{function-centered} abstraction. In this classical view, a service is primarily characterized by its interface, input-output contract, quality attributes, and externally invoked execution logic. While this paradigm has successfully powered the digital economy, it inherently assumes that the service object is a passive, stateless, or deterministically stateful endpoint whose behavior is fully specified at design time.

\begin{figure}[t]
    \centering
    \includegraphics[width=0.9\linewidth]{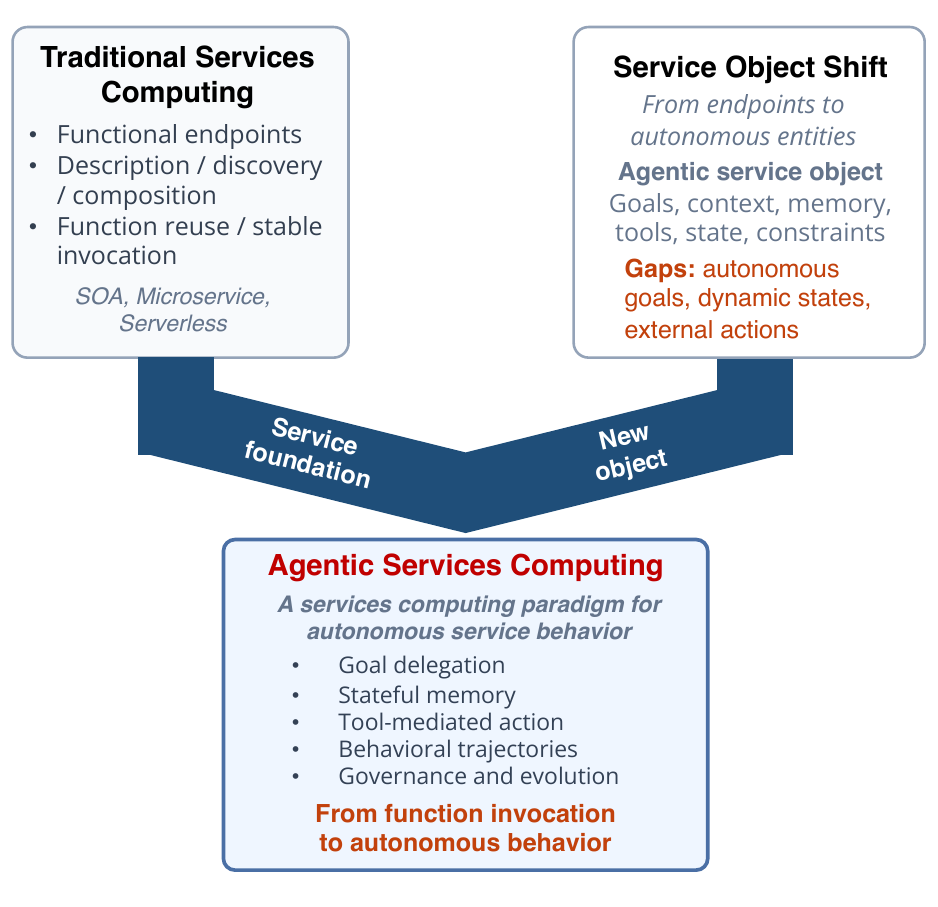}
    \caption{Our definition of Agentic Services Computing is motivated by the ontological shift of the service object from functional endpoints to autonomous service entities. ASC extends services computing from managing functional invocations to engineering and governing goal-driven autonomous service behavior.}
    \label{fig:asc_definition}
\end{figure}

The rapid ascent of large language model (LLM)-based agents is precipitating an ontological shift in the service object itself. Modern agents are no longer mere functions to be invoked; they are autonomous entities capable of interpreting high-level goals, maintaining long-term context and memory, orchestrating tools, collaborating with peers, and dynamically revising execution trajectories based on environmental feedback~\cite{yao2023react,shinn2023reflexion,wang2024voyager,zhou2024webarena,koh2024visualwebarena,wu2024autogen}. In critical domains such as cloud incident diagnosis, software maintenance, enterprise workflow automation, privacy auditing, and personalized recommendation, these \textit{agentic services} are increasingly deployed as operational entities that act on behalf of users and organizations~\cite{xu2025openrca,wang2024rcagent,zhang2025tamo,mao2025agentic,chen2025stratus,chen2025locagent,guo2025repoaudit,zhang2026priagent,liu2026mars}. Their value proposition lies not in executing predefined operations, but in generating outcomes through autonomous service behavior that interprets delegated intent, leverages contextual knowledge, exposes observable intermediate states, and delivers results under practical constraints.


\begin{figure*}[t]
    \centering
    \includegraphics[width=\textwidth]{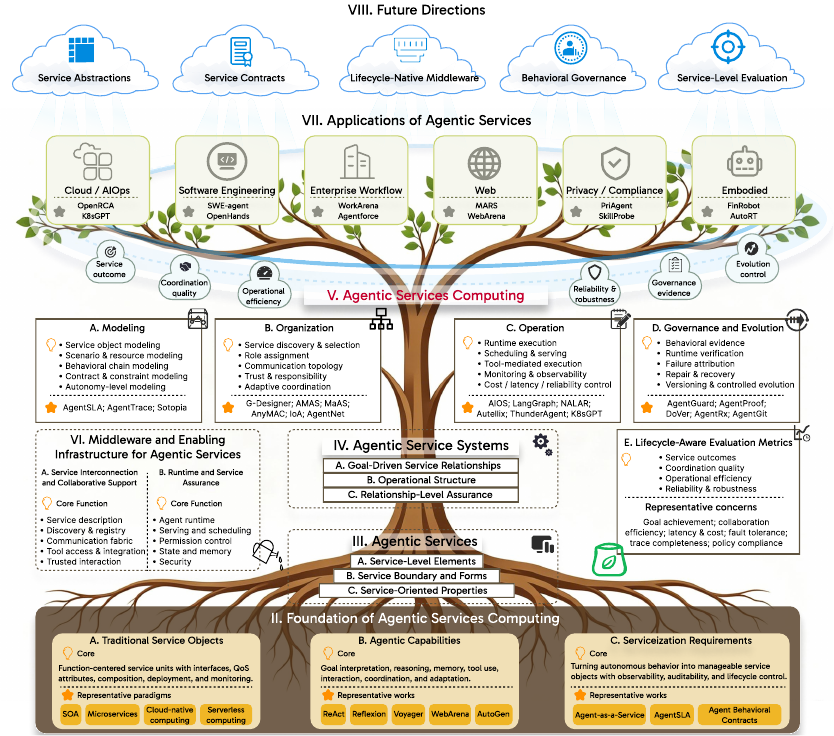}
    \caption{Organization of the article. The survey proceeds from conceptual grounding to lifecycle engineering and then to ecosystem outlook, with each part mapped to the corresponding sections.}
    \label{fig:section_structure}
\end{figure*}

This fundamental shift exposes a critical gap in existing service abstractions. Traditional services computing provides mature foundations for description, discovery, composition, delivery, QoS management, and governance. However, these foundations are predicated on stable interfaces and predictable execution paths. Agentic service behavior, by contrast, involves goals, evolving contexts, tool-mediated action spaces, non-deterministic behavioral trajectories, permission boundaries, side effects, and accountability evidence. When service fulfillment is mediated by dynamic planning and external actions, failures transcend simple output errors to become service-level failures affecting execution integrity and downstream consequences~\cite{xu2025reducing,liu2026agenthallu,zhang2026litmus}. The central challenge is therefore not merely how to invoke agents as intelligent functions, but how to model, compose, operate, audit, and evolve \textit{autonomous service behavior} within a rigorous service-object framework.

\begin{table*}[t]
    \centering
    \caption{A comparative analysis of representative perspectives and systems related to agentic services. This work takes services computing as the primary lens and organizes related studies and systems around the lifecycle of autonomous service entities.}
    \label{tab:survey_perspectives}
    \setlength{\belowcaptionskip}{8pt}
    
    \setlength{\tabcolsep}{2.6pt}
    \renewcommand{\theadfont}{\footnotesize\bfseries}
    \renewcommand{\theadgape}{}
    \renewcommand{\arraystretch}{1.05}
    
    \begin{tabular}{@{}c ccccccc @{}}
        \toprule
        \textbf{Work} & 
        \makecell{\textbf{Trust} \\ \textbf{\& Alignment} \\
        \scriptsize\cite{zeng2025multilevelvaluealignmentagentic,huang2024survey} \\
        \scriptsize\cite{ferdaus2024towards,donta2025socio}} &
        \makecell{\textbf{Agent} \\ \textbf{Architectures} \\
        \scriptsize\cite{wang2024survey,xi2025rise} \\
        \scriptsize\cite{krishnan2025ai}} &
        \makecell{\textbf{Domain} \\ \textbf{Applications} \\
        \scriptsize\cite{zeng2023large,duan2022survey,yang2023llm4drive} \\
        \scriptsize\cite{thirunavukarasu2023large,du2021vision,liu2024large,he2025llm}} &
        \makecell{\textbf{Multi-Agent} \\ \textbf{\& SE} \\
        \scriptsize\cite{li2024survey,guo2024large} \\
        \scriptsize\cite{he2025llm,liu2024large}} &
        \makecell{\textbf{Agentic Service} \\ \textbf{Ecosystems} \\
        \scriptsize\cite{zhang2025survey}} &
        \makecell{\textbf{Operational} \\ \textbf{Agent Platforms} \\
        \scriptsize\cite{harness2026agents,openclaw2026homepage}} &
        \makecell{\textbf{ASC (Ours)}} \\
        \midrule
        \textbf{LLM Agents} & \ding{51} & \ding{51} & \ding{51} & \ding{51} & \ding{55} & \ding{51} & \textbf{\ding{51}} \\
        \textbf{Agent Collaboration} & \ding{51} & \ding{51} & \ding{55} & \ding{51} & \ding{51} & \ding{55} & \textbf{\ding{51}} \\
        \textbf{Service Abstraction} & \ding{55} & \ding{55} & \ding{55} & \ding{55} & \ding{51} & \ding{51} & \textbf{\ding{51}} \\
        \textbf{Object Shift} & \ding{55} & \ding{55} & \ding{55} & \ding{55} & \ding{55} & \ding{51} & \textbf{\ding{51}} \\
        \textbf{Service Lifecycle} & \ding{55} & \ding{55} & \ding{55} & \ding{55} & \ding{55} & \ding{55} & \textbf{\ding{51}} \\
        \textbf{Governance \& Metrics} & \ding{51} & \ding{51} & \ding{51} & \ding{51} & \ding{51} & \ding{51} & \textbf{\ding{51}} \\
        \bottomrule
    \end{tabular}
\end{table*}

Research on LLM agents has expanded rapidly across architectures, multi-agent collaboration, safety, observability, and governance~\cite{wang2024survey,xi2025rise,wooldridge2009introduction,jennings2001agent,guo2024large,li2024survey,yang2025agentic,laju2026nalar,kang2026thunderagent,dong2024agentops,alsayyad2026agenttrace,zheng2025agentsight,koohestani2025agentguard,xavier2026agentproof,mavravcic2025policy}. Concurrently, operational platforms have begun to treat agents as managed execution units. For instance, \textit{Harness Agents} operationalize AI agents as workers within DevSecOps pipelines, leveraging pipeline context, secrets, access control, and policy gates to support software delivery~\cite{harness2026agents}. Similarly, \textit{OpenClaw} embeds autonomous personal agents into local and messaging environments to execute tasks across external tools~\cite{openclaw2026homepage}. These systems demonstrate that autonomous agents are transitioning from interactive assistants to embedded operational entities with governance requirements. However, their abstractions remain tightly coupled to specific engineering substrates or personal-assistant paradigms. As summarized in Table~\ref{tab:survey_perspectives}, while existing studies and systems provide essential foundations for agent capabilities, collaboration, and platform-level governance, a service-centered roadmap that explains how autonomous behavior itself becomes a describable, composable, operable, governable, and evolvable \textit{service object} remains underdeveloped.

Against this background, in this work, we introduce \textit{Agentic Services Computing} (ASC), a new paradigm that extends services computing from managing reusable functional endpoints to engineering and governing autonomous service entities. ASC treats \textit{agentic services} as first-class service objects: service-oriented autonomous agents that encapsulate delegated goals, capabilities, context, memory, tool access, behavioral constraints, operational state, quality expectations, and governance interfaces. As illustrated in Figure~\ref{fig:asc_definition}, ASC preserves the core concerns of services computing, i.e., description, composition, delivery, and governance, while shifting the engineering target from function invocation to autonomous service behavior.
This article develops ASC as a comprehensive, service-centered research roadmap. It first establishes the conceptual foundation by clarifying the service-object shift and defining agentic services. It then examines agentic service systems, where autonomous entities form goal-driven relationships through roles, protocols, shared context, and governance boundaries. The roadmap is structured around four lifecycle stages: (i) \textit{modeling}, which extends service description to encompass goals, capabilities, contexts, tools, contracts, and risk boundaries; (ii) \textit{organization}, which evolves service composition from predefined workflows to adaptive multi-agent coordination; (iii) \textit{operation}, which manages goal-driven service fulfillment under runtime constraints; and (iv) \textit{governance and evolution}, which enables behavioral auditing, accountability, controlled change, and continuous improvement. The article further discusses enabling middleware, lifecycle-aware metrics, representative applications, and open challenges.

The remainder of this article is organized as shown in Figure~\ref{fig:section_structure}. Section~\ref{sec:foundation} provides conceptual grounding by analyzing traditional service objects, agentic capabilities, and serviceization requirements. Section~\ref{sec:agentic-services} defines agentic services through service-level elements, boundaries, and properties. Section~\ref{sec:agentic-service-systems} discusses agentic service systems via goal-driven relationships and relationship-level assurance. Section~\ref{sec:asc} presents the ASC lifecycle engineering framework, including modeling, organization, operation, governance, evolution, and metrics. Section~\ref{sec:middleware} covers interconnection, collaboration, runtime, and assurance middleware. Section~\ref{sec:applications} reviews representative application evidence, Section~\ref{sec:future} identifies future research directions, and Section~\ref{sec:conclusion} concludes the article.

\section{Foundation}\label{sec:foundation}


ASC is motivated by a shift in the object managed by services computing. Traditional paradigms have primarily engineered reusable software functions, whereas emerging LLM-based agents introduce service behavior that is goal-driven, stateful, tool-mediated, collaborative, and capable of producing external side effects. The foundation of ASC is therefore not merely the integration of agent capabilities into service systems, but the \textit{serviceization} of autonomous agent behavior, through which such behavior becomes describable, composable, operable, monitorable, governable, and evolvable as a first-class service object. This section establishes the necessity of ASC through three interconnected arguments: the limitations of traditional service objects, the ontological distinctiveness of agentic capabilities, and the engineering requirements of serviceization.

\subsection{Traditional Service Objects and Their Limits}
Services computing studies how computational capabilities are abstracted, exposed, and managed as services. From service-oriented architectures~\cite{papazoglou2003service,papazoglou2007service,bichier2006service} to microservices, cloud-native, and serverless paradigms~\cite{newman2021building,baldini2017serverless,deng2024cloud}, the discipline has matured around a function-centered abstraction. A traditional service is specified by its interface, input-output contract, quality attributes, deployment unit, and operational state. It can be discovered, selected, composed, scaled, and monitored, but its behavior remains largely determined by predefined implementation logic and externally invoked execution.

This abstraction is effective for stable functional endpoints but becomes insufficient when the service object exhibits autonomous, context-sensitive, and tool-mediated behavior with external side effects. Traditional service models assume deterministic or bounded-state execution paths, making them ill-equipped to handle non-deterministic behavioral trajectories, dynamic goal interpretation, or accountability for unintended consequences. The challenge thus transcends exposing and composing functions; it lies in the serviceization of autonomous behavior: making such behavior describable, controllable, operable, and governable across its lifecycle. This limitation constitutes the primary motivation for ASC.

\subsection{Agentic Capabilities as the New Service Substrate}
Autonomous agents provide the capability basis for this ontological shift. Classical multi-agent systems established agents as autonomous entities that perceive, reason, communicate, and coordinate in distributed settings~\cite{wooldridge2009introduction,jennings2001agent}. However, their reliance on hand-crafted knowledge and domain-specific reasoning limited adaptability in open service environments. Recent LLM-based agents overcome these constraints through natural language goal understanding, contextual memory, tool use, feedback-driven reasoning, and interaction with open digital ecosystems. Foundational works such as ReAct~\cite{yao2023react}, Reflexion~\cite{shinn2023reflexion}, Voyager~\cite{wang2024voyager}, WebArena~\cite{zhou2024webarena}, VisualWebArena~\cite{koh2024visualwebarena}, and AutoGen~\cite{wu2024autogen} demonstrate interleaved reasoning and acting, self-improvement, skill acquisition, realistic environment interaction, and multi-agent collaboration.

From a services computing perspective, these capabilities matter because they redefine what a service is expected to do. An agentic service is no longer confined to executing predefined functions; it can interpret delegated goals, orchestrate context-aware tool usage, coordinate with peers, and act on behalf of users or organizations. Empirical evidence from cloud incident diagnosis~\cite{xu2025openrca,wang2024rcagent}, software maintenance~\cite{guo2025repoaudit}, enterprise workflow automation~\cite{zhang2025tamo,chen2025stratus}, privacy auditing~\cite{zhang2026priagent}, and service recommendation~\cite{liu2026mars} confirms a shift from isolated capability evaluation toward delegated service fulfillment in real-world settings. Yet capability alone does not constitute a service object. For services computing, autonomous behavior must be encapsulated within explicit service boundaries before it can be reused, composed, monitored, audited, and evolved. This gap necessitates the serviceization process.

\subsection{Serviceization}
Serviceization is the systematic process through which autonomous agent behavior is transformed into a manageable service entity. A general-purpose agent may reason, plan, and interact with environments, but production service ecosystems require stable identity, explicit description, contractual constraints, permission boundaries, runtime observability, audit mechanisms, and lifecycle management. These properties distinguish an agentic service from an ordinary agent: traditional services are manageable but function-centric; ordinary agents are autonomous but lack service-level manageability; agentic services unify both by making autonomous behavior itself a service object.

Serviceization extends core services computing constructs. \textit{Description} must capture not only functional interfaces and QoS attributes, but also acceptable goals, claimed capabilities, contextual dependencies, tool-mediated action spaces, operational constraints, and observable behavioral traces. \textit{Composition} evolves from connecting predefined functions to organizing autonomous entities, including selecting agents, assigning roles, establishing protocols, coordinating tool use, maintaining task state, and controlling error propagation. \textit{Governance} must address behavioral auditing, accountability attribution, and controlled evolution under organizational and regulatory requirements.

Recent work illustrates partial realizations of these requirements. Agent-as-a-Service frameworks organize agents through registration, discovery, interoperability protocols, and execution graphs~\cite{zhu2025agent}. AgentSLA extends service-level agreements to model quality attributes and objectives for AI agents~\cite{jouneaux2025agentsla}, while Agent Behavioral Contracts specify preconditions, invariants, governance policies, and recovery mechanisms as runtime-enforceable components~\cite{bhardwaj2026agent}. These efforts demonstrate that autonomous behavior must be anchored to identity, quality expectations, behavioral boundaries, runtime evidence, and governance mechanisms to become a dependable service entity. Collectively, these requirements establish the foundation of ASC: agentic capabilities explain why autonomous service behavior is possible, while serviceization explains how such behavior becomes manageable within services computing. Through identity, description, contracts, permissions, observability, audit, and lifecycle management, autonomous behavior is elevated from an agent capability to a service-level engineering object.

\section{Agentic Services}\label{sec:agentic-services}


Agentic services are the fundamental service objects managed by ASC. They are introduced to capture a class of entities that cannot be fully described as callable functions nor treated as ordinary agents. A traditional service is exposed through a stable interface and invoked to perform predefined operations; an ordinary agent may reason, plan, and interact with environments but lacks service identity, constraints, runtime evidence, and lifecycle governance. An agentic service bridges this gap by transforming autonomous agent behavior into a manageable service entity.

\begin{definition}[Agentic Service]
An agentic service is an autonomous agent exposed and managed as a service entity, whose delegated or authorized goals and goal-driven, stateful, tool-mediated, and interactive behavior are represented through service-level artifacts for description, discovery, composition, execution, monitoring, auditing, and evolution under explicit constraints.
\end{definition}

\begin{figure}[t]
    \centering
    \includegraphics[width=\linewidth]{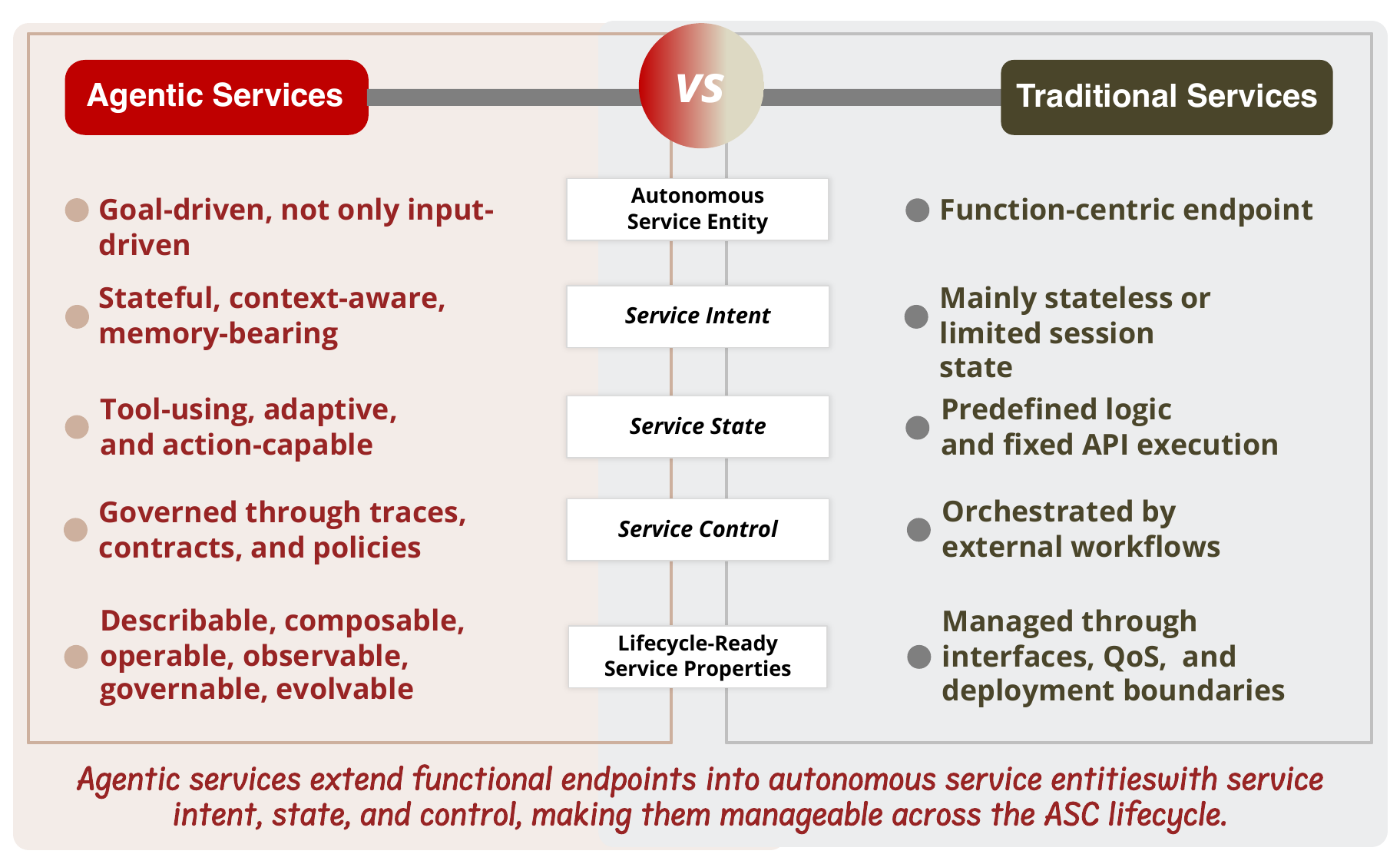}
    \caption{Comparison between traditional services and agentic services. Agentic services extend function-centric endpoints into autonomous service entities whose behavior is goal-driven, stateful, tool-mediated, and managed through service-level constraints.}
    \label{fig:compare}
\end{figure}

This definition comprises two orthogonal dimensions. The first is \textit{autonomous service behavior}: the capacity to accept delegated goals, interpret them in context, select tools beyond fixed operations, interact with users or peers, and adapt execution based on feedback. Recent LLM-based agents demonstrate this in cloud operations, software engineering, privacy auditing, and service recommendation~\cite{yao2023react,wang2024rcagent,xu2025openrca,zhang2025tamo,chen2025locagent,zhang2026priagent,liu2026mars}. In proactive settings, goals may also be triggered by authorized events, monitoring signals, or environmental observations~\cite{lu2025proactive,deng2024towards}. The second dimension is \textit{service-level manageability}: autonomous behavior becomes a service object only when anchored to stable identity, capability claims, permissions, quality expectations, runtime traces, intervention points, and governance artifacts. This duality is reflected in emerging frameworks for agent-level SLAs, behavioral contracts, policy enforcement, and interoperability protocols~\cite{zhu2025agent,jouneaux2025agentsla,bhardwaj2026agent,mavravcic2025policy,yang2025agentic,mcp,a2a}. As illustrated in Figure~\ref{fig:compare}, agentic services preserve service exposure while extending the managed object from functional endpoints to autonomous entities governed by service-level constraints.

\subsection{Service-Level Elements}
\begin{figure}[t]
    \centering
    \includegraphics[width=\linewidth]{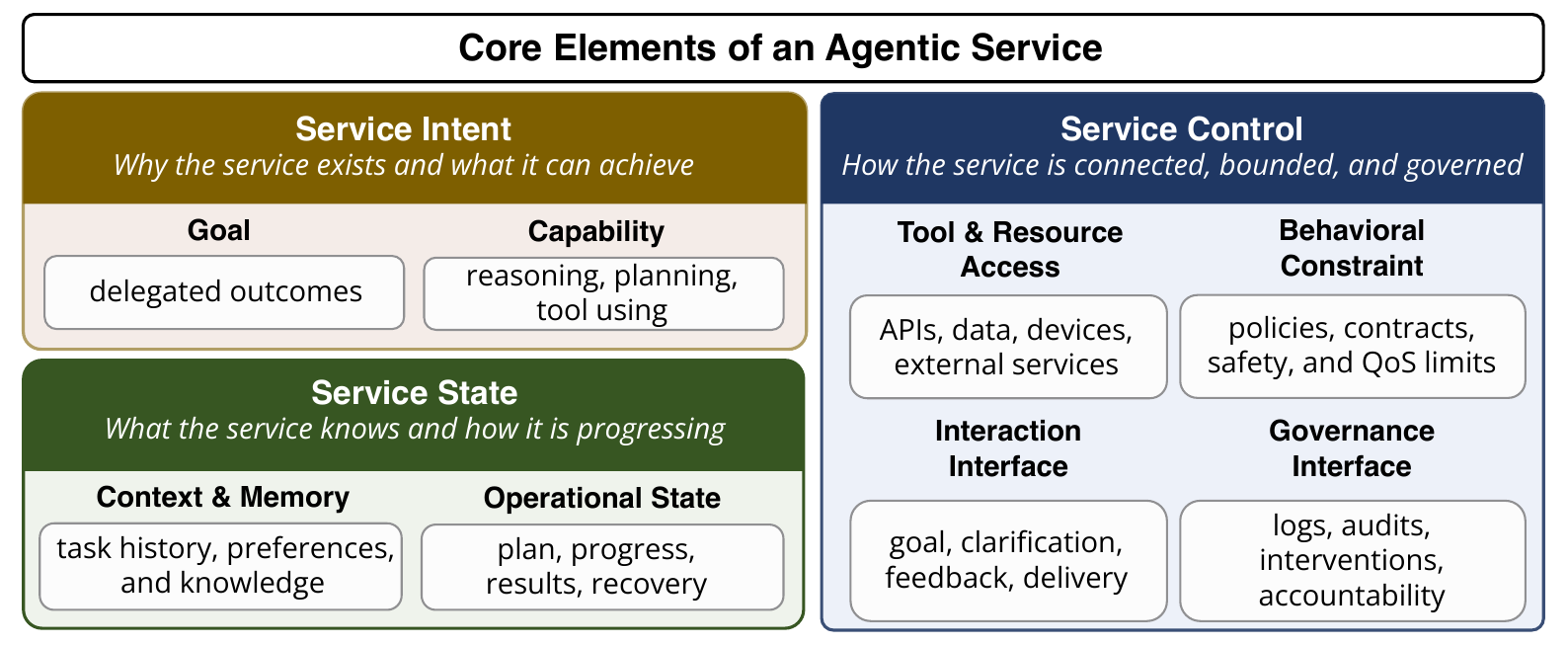}
    \caption{Core elements of an agentic service. Service intent specifies delegated goals and capabilities; service state captures context, memory, and execution progress; service control defines connection, boundaries, and governance interfaces.}
    \label{fig:core_elements}
\end{figure}

To make autonomous behavior manageable, ASC represents an agentic service through three core elements, i.e., intent, state, and control, that specify what is delegated, what continuity is carried, and how behavior is bounded. These elements extend traditional service descriptions from interfaces and QoS attributes to the conditions under which autonomous behavior is executed, constrained, observed, and improved (see Figure~\ref{fig:core_elements}).

\textit{Service intent} specifies what is delegated and what the service claims to provide. Unlike traditional invocation where callers select operations with fixed inputs, agentic invocation delegates intended outcomes that must be interpreted, decomposed, and evaluated during execution. The goal anchors planning, tool selection, exception handling, and completion assessment. Capability describes what the service can achieve under stated assumptions, encompassing reasoning, tool use, domain knowledge, collaboration, and human interaction. Its description must include task scope, applicable contexts, resource requirements, quality objectives, autonomy level, oversight needs, and known limitations~\cite{jouneaux2025agentsla,zhu2025agent,porter2025insyte}. Intent thus serves as a service-level claim about delegated goals and capability boundaries.

\textit{Service state} captures continuity within and across executions. Agentic services rely on task context, user preferences, historical interactions, retrieved knowledge, and long-term memory to support multi-turn dialogue, long-horizon workflows, and personalized delivery. However, statefulness introduces risks such as stale context, memory contamination, privacy leakage, and cross-task interference~\cite{zhang2026memskill,huang2026ama,wu2025evolver}. Operational state tracks execution progress through reasoning, planning, tool invocation, and collaboration, affecting future actions and requiring checkpointing, rollback, or recovery mechanisms~\cite{zhou2025shielda,li2025agentgit}. From a service perspective, state must be characterized by its scope, source, update rules, retention policy, visibility, access controls, recovery semantics, and evolution history.

\textit{Service control} defines how autonomous behavior is connected, bounded, and governed. Tool and resource access links the service to external systems (APIs, databases, cloud platforms, devices), and because tool invocation produces side effects, it must specify available tools, schemas, authentication, permission boundaries, rate limits, failure modes, validation rules, and compensation strategies~\cite{yao2023react,wang2024rcagent,mao2025agentic,mcp,a2a}. Behavioral constraints encode permissions, preconditions, invariants, prohibited actions, escalation conditions, safety requirements, and compliance rules~\cite{bhardwaj2026agent,mavravcic2025policy,paduraru2026trace,kaptein2026runtime}. Interaction interfaces support goal submission, clarification, progress reporting, interruption, and escalation. Governance interfaces expose traces, logs, policies, versions, audit records, intervention points, and evidence artifacts, enabling monitoring, auditing, and attribution of non-deterministic trajectories and side effects~\cite{dong2024agentops,alsayyad2026agenttrace,zheng2025agentsight,koohestani2025agentguard,xavier2026agentproof}.

Together, intent, state, and control form the internal structure of an agentic service: intent defines delegation and capability claims, state maintains continuity and enables recovery, and control enforces boundaries and ensures governability. These elements render autonomous behavior explicit enough for services computing management.

\subsection{Service Boundary and Forms}
While service-level elements describe internal composition, the service boundary specifies when an autonomous agent qualifies as a service object. An agent becomes an agentic service when its behavior is exposed through stable identity, explicit description, controlled tool access, observable execution, operational state, quality expectations, and governable lifecycle interfaces. In Agent-as-a-Service settings, this boundary materializes via registration, discovery, self-describing contracts, task routing, scoped execution, and cross-domain collaboration~\cite{su2025role,kang2026openaaas}. This criterion distinguishes deployable service entities from agents that merely demonstrate task-solving ability.

The boundary is especially critical for proactive behavior, where services act before explicit user instruction. Such behavior must be tied to authorized triggers, contextual validity, cost/risk constraints, audit evidence, and human-centered interaction requirements (timing, adaptivity, non-intrusiveness)~\cite{deng2024towards,lu2025proactive}. The value of an agentic service thus depends not only on autonomous capability but also on clear service scope, expected outcomes, responsibility attribution, and governance interfaces.

Agentic services manifest in recurring, often overlapping forms: \textit{task-oriented} services fulfill delegated goals in bounded domains; \textit{tool-augmented} services combine reasoning with controlled external access; \textit{workflow} services execute long-horizon processes with planning, state management, and recovery; \textit{composite} services coordinate specialized agents via roles, protocols, and shared context; \textit{governed} services embed policies, contracts, and runtime verification~\cite{wang2024rcagent,xu2025openrca,chen2025locagent,guo2025repoaudit,zhang2026priagent,liu2026mars,wu2024autogen,mavravcic2025policy,bhardwaj2026agent}. These are implementation patterns rather than disjoint categories; practical services (e.g., cloud diagnosis, code auditing) typically combine multiple forms. What unifies them is the serviceization of autonomous behavior: goals, capabilities, context, tools, constraints, state, quality expectations, traces, and governance interfaces become integral parts of the service object, keeping ASC focused on entities that can be described, delivered, monitored, audited, and evolved within a service ecosystem.

\subsection{Service-Oriented Properties}
\begin{figure}[t]
    \centering
    \includegraphics[width=\linewidth]{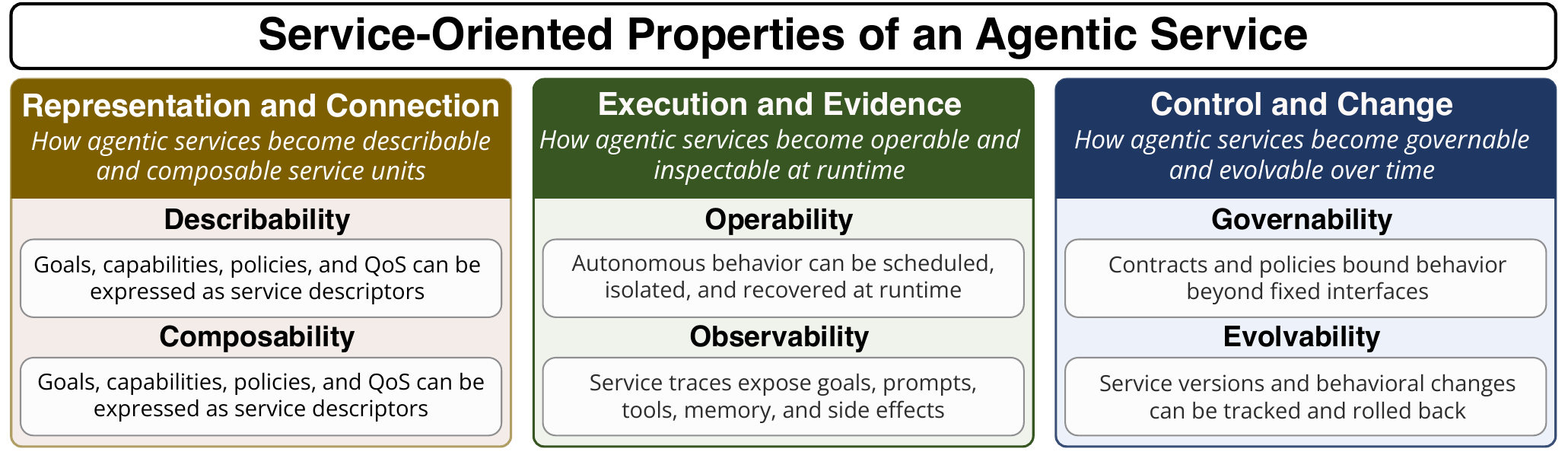}
    \caption{Service-oriented properties of an agentic service. These properties enable agentic services to function as describable/composable units, operable/inspectable runtime entities, and governable/evolvable service objects.}
    \label{fig:service_properties}
\end{figure}

Once defined, represented, and bounded, agentic services are characterized by properties that enable lifecycle management. These properties explain how they enter the ASC lifecycle as objects for modeling, organization, operation, governance, and evolution (Figure~\ref{fig:service_properties}).

\textit{Representation and connection} concern describability and composability. Describability ensures that goals, capabilities, context/memory scope, tool permissions, state, constraints, quality expectations, and governance interfaces are represented in forms supporting discovery, selection, composition, operation, and governance~\cite{jouneaux2025agentsla,mavravcic2025policy,bhardwaj2026agent}. Composability enables participation in larger service structures through role compatibility, communication protocols, shared context, tool dependencies, state handoff, policy alignment, and recovery semantics~\cite{wu2024autogen,gao2026rcaflow,zhang2026priagent,liu2026mars}.

\textit{Execution and evidence} concern operability and observability. Operability means the service can be executed, scheduled, isolated, scaled, recovered, and delivered under latency, cost, reliability, and resource constraints~\cite{mei2025aios,laju2026nalar,2R,kang2026thunderagent}. Observability ensures that goals, prompts, reasoning steps, tool calls, messages, memory updates, workflow states, system calls, and interventions are recorded and verifiable during and after execution~\cite{dong2024agentops,alsayyad2026agenttrace,zheng2025agentsight}. These are essential because failures may manifest as incorrect plans, invalid tool parameters, ungrounded reports, unsafe actions, or inconsistent state—not just output errors.

\textit{Control and change} concern governability and evolvability. Governability ensures permissions, policies, contracts, audit trails, intervention mechanisms, rollback records, and responsibility boundaries are defined and enforced throughout execution~\cite{mavravcic2025policy,bhardwaj2026agent,koohestani2025agentguard,xavier2026agentproof}. Evolvability allows updates to prompts, tools, memories, policies, roles, descriptions, and deployments while preserving version records, evaluation evidence, compatibility constraints, and rollback paths. This is necessary because agentic services learn, adapt, and change behavior through model, tool, memory, or policy updates.

These properties do not make an agentic service a stronger agent in a general sense; they make it a service object whose autonomous behavior can be described before execution, connected with other services, delivered under constraints, inspected through evidence, constrained by governance, and improved over time. Thus, agentic services provide the object-level foundation for ASC.

\section{Agentic Service Systems}\label{sec:agentic-service-systems}


An agentic service system represents the system-level manifestation of ASC. Once autonomous agents are exposed as service entities, the fundamental system challenge shifts from connecting heterogeneous functions to governing service relationships among goal-driven, stateful, and tool-mediated entities. Traditional service systems organize relationships through interfaces, workflows, message exchanges, deployment dependencies, and QoS constraints~\cite{papazoglou2003service,papazoglou2007service,newman2021building,baldini2017serverless,deng2024cloud}. Agentic service systems preserve these foundational relations but introduce new relationship semantics centered on goal delegation, role coordination, shared context, tool-mediated action, and governance responsibility.

\begin{definition}[Agentic Service System]
An agentic service system is a service system composed of agentic services and their supporting users, tools, resources, data, infrastructures, and governance mechanisms, wherein autonomous service entities form dynamic service relationships around delegated goals, coordinated roles, shared contexts, controlled tool access, operational constraints, and responsibility boundaries to deliver auditable service outcomes in open and evolving environments.
\end{definition}

Critically, an agentic service system is bounded by the \textit{service fulfillment process} rather than by the deployment perimeter of a single software stack. Existing APIs, microservices, cloud platforms, and enterprise data systems remain the substrate for exposing functions and resources. Agentic services extend this substrate by interpreting goals, carrying task context, invoking tools, coordinating with peers, and producing evidence for monitoring and accountability. The system boundary thus encompasses all participants, resources, runtime supports, and governance mechanisms that jointly shape service outcomes under shared operational constraints.

\subsection{Goal-Driven Service Relationships}
Traditional service systems succeed by transforming heterogeneous software into manageable units, enabling reuse through description, interoperability through standard interfaces, composition through workflows, and dependability through QoS management~\cite{ngan2013semantic,andrews2003business,mayer2008model,zeng2003quality,jatoth2015computational,ghafouri2020survey}. This abstraction fits tasks whose fulfillment can be specified as predefined function composition. However, it becomes insufficient when service fulfillment requires autonomous interpretation, adaptation, and coordination beyond static invocation chains.

\begin{figure}[t]
    \centering
    \includegraphics[width=\linewidth]{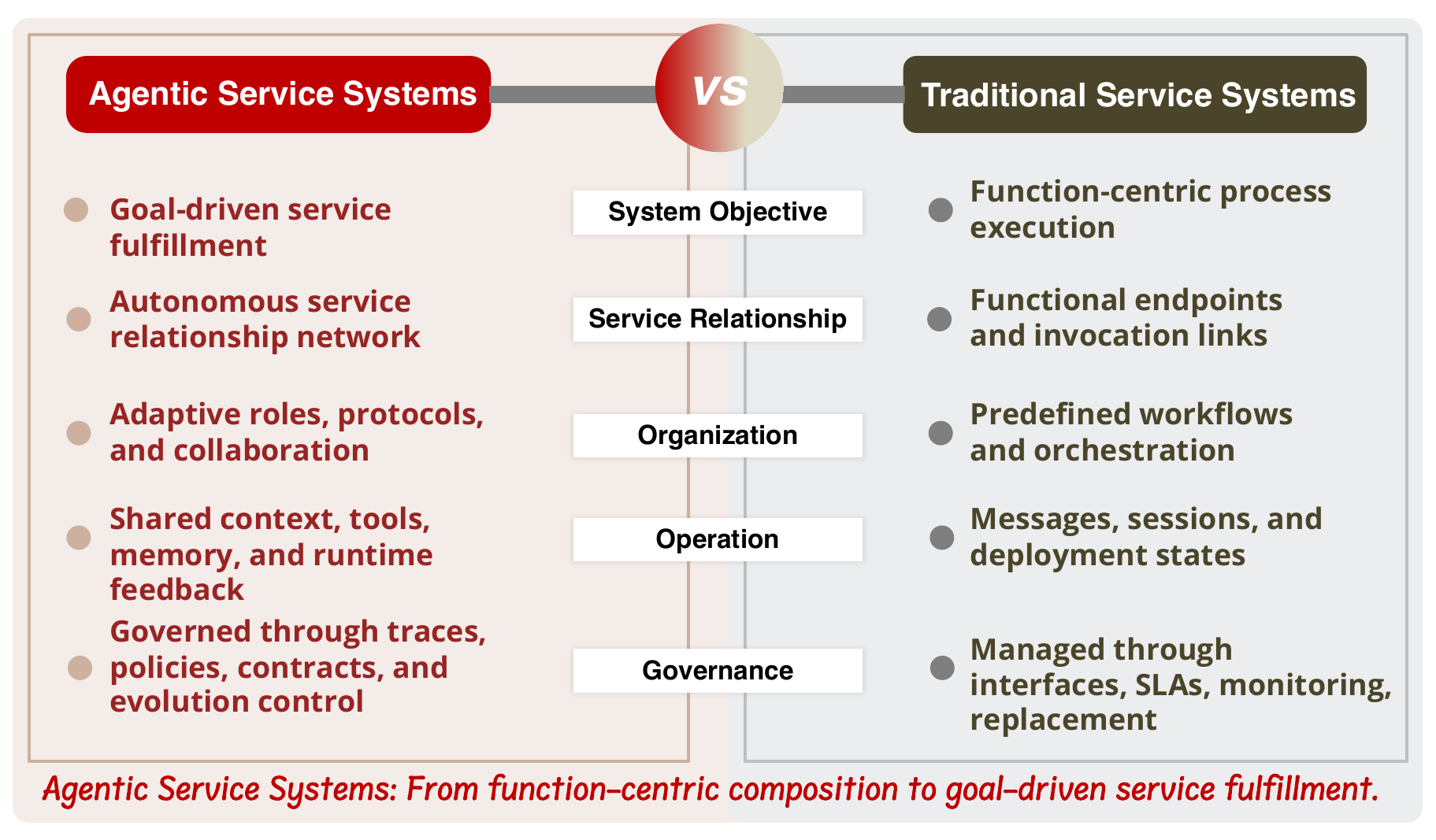}
    \caption{Relationship-level comparison between traditional and agentic service systems. Agentic service systems extend function-centric endpoints into goal-driven service fulfillment through goal delegation, role coordination, shared context, tool-mediated action, and governance responsibility as first-class relations.}
    \label{fig:system_comparison}
\end{figure}

Figure~\ref{fig:system_comparison} illustrates this relationship-level shift. While traditional systems organize stable endpoints via invocation and deployment relations, agentic service systems add {goal delegation}, {role coordination}, {shared context}, {tool-mediated action}, and {governance responsibility} as first-class system-level relations. These are not merely additional connections; they represent a semantic transformation of service interaction. In cloud incident response, for example, mitigation depends on interpreting alerts, logs, metrics, traces, and dependency graphs before selecting actions~\cite{wang2024rcagent,xu2025openrca,zhang2025tamo,chen2025stratus,pei2025flow}. In repository-level code auditing, fulfillment requires navigating code, reasoning about dependencies, executing tools, and synthesizing findings~\cite{chen2025locagent,guo2025repoaudit,he2025llm}. In service recommendation, agents must bridge user requirements and API descriptions while respecting composition patterns and runtime constraints~\cite{liu2026mars,wang2024macrec,zu2026think}. These remain service tasks, but their fulfillment relies on autonomous behavior distributed across dynamically related entities.

Consequently, agentic service systems shift from function-centric composition to \textit{goal-driven service fulfillment}. A delegated goal is interpreted, decomposed, routed, negotiated, and revised across roles, contexts, tools, and intermediate results. Service fulfillment is organized through relationships that explicitly connect goals, capabilities, roles, shared states, external actions, and responsibility boundaries, making the system's adaptive behavior itself a manageable engineering artifact.

\subsection{Operational Structure}
\begin{figure}[t]
    \centering
    \includegraphics[width=\linewidth]{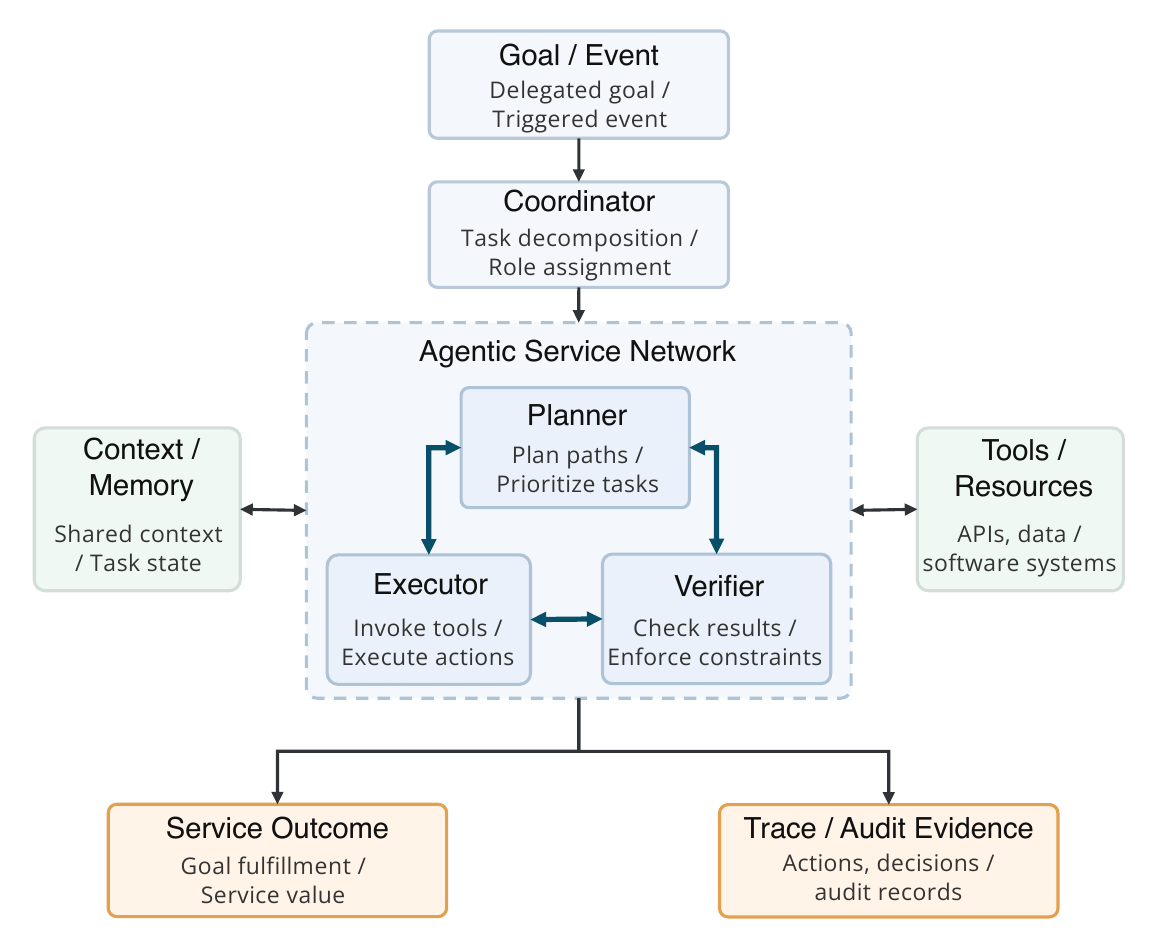}
    \caption{Operational structure of an agentic service system. A delegated or triggered goal is coordinated through planning, execution, and verification, supported by context, memory, tools, and resources, and delivered as a service outcome together with trace and audit evidence.}
    \label{fig:system_capability}
\end{figure}

To transform delegated or triggered goals into service outcomes under runtime and governance constraints, an agentic service system requires an operational structure that serves as an \textit{auditable service fulfillment chain}. As summarized in Figure~\ref{fig:system_capability}, a goal or event enters a coordinator that decomposes the objective, assigns roles, and orchestrates execution. The agentic service network then plans execution paths, invokes tools and resources, validates results against constraints, and updates shared context or task state. Crucially, unlike ordinary agent workflows, this structure must produce not only service outcomes but also trace and audit evidence for monitoring, accountability, and subsequent governance.

This fulfillment-chain view explicitly links goals to coordinated execution, context/memory to task continuity, tools/resources to controlled external action, and audit evidence to governance. Multi-agent frameworks and service-oriented agent networks provide foundational mechanisms for roles, protocols, tool interfaces, shared contexts, and runtime coordination~\cite{wu2024autogen,hong2024metagpt,liao2025agentmaster,zhu2025agent}. Adaptive topologies, optimizable agent graphs, dynamic communication, and task scheduling demonstrate how such structures evolve with task semantics, input characteristics, and resource constraints~\cite{AMAS2025,MaAS2025,wang2025dynamic,yu2025dyntaskmas,zhao2025cola}. In practice, cloud operations, software engineering, and service recommendation systems already integrate agents with data, tools, workflows, human knowledge, and operational policies~\cite{wang2024rcagent,chen2025stratus,guo2025repoaudit,liu2026mars}, validating the feasibility of this operational structure.

\subsection{Relationship-Level Assurance}
\begin{figure}[t]
    \centering
    \includegraphics[width=0.67\linewidth]{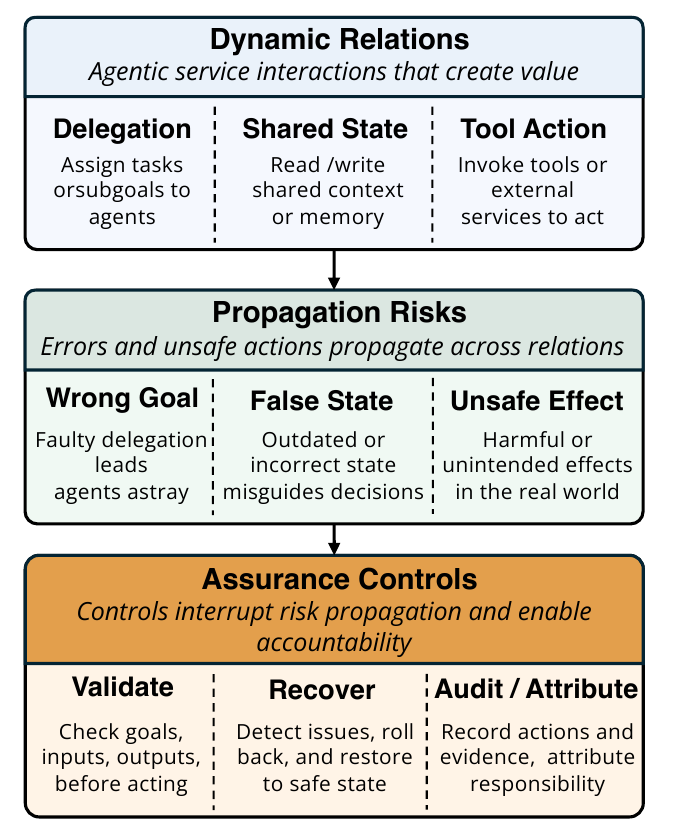}
    \caption{Relationship-level assurance in agentic service systems. Dynamic relations such as delegation, shared state, and tool-mediated action propagate wrong subgoals, false states, and unsafe side effects, requiring validation, recovery, auditing, and attribution controls.}
    \label{fig:system_character_risk}
\end{figure}

Because agentic service systems rely on dynamic service relationships, assurance must be elevated to a \textit{relationship-level concern}. Delegation, shared state, and tool-mediated action enable distributed service fulfillment but also create systemic pathways for error propagation and unsafe behavior. Faulty delegation may yield incorrect subgoals; stale or contaminated shared state may misguide downstream decisions; unsafe tool actions may produce irreversible external side effects. Relationship-level assurance therefore requires controls that validate goals, inputs, outputs, and actions \textit{before} execution; recover service state \textit{after} failures; and preserve audit evidence \textit{throughout} for responsibility attribution (Figure~\ref{fig:system_character_risk}).

These risks transcend final-answer errors. Execution hallucination exemplifies this: an agent may report progress, select actions, update memory, or produce intermediate claims ungrounded in reality. Once such claims enter messages, shared memory, delegated subtasks, or workflow states, they propagate through the service relationship graph and complicate responsibility attribution~\cite{liu2026agenthallu}. Tool hallucination and execution validation studies confirm that assurance must verify consistency among claims, actions, tool parameters, and environment states~\cite{xu2025reducing,zhang2026litmus}.

Additional risks stem directly from service relationships. Dynamic collaboration increases communication overhead, duplicated reasoning, and coordination latency. Shared context and long-term memory risk stale knowledge, privacy leakage, semantic drift, and reinforcement of early errors~\cite{lu2026auditing,suresh2026noise}. Open tool access introduces permission and security vulnerabilities. Multiple autonomous services may cause policy conflicts, ambiguous responsibility boundaries, and intractable failure attribution. Empirical studies on multi-agent workflow failures reveal common patterns: weak task verification, missing history, task derailment, reasoning-action mismatch, premature termination, and incomplete validation~\cite{WhyMASFail,qi2026beyond,MultiAgentBench}. Communication attacks, topology-aware exploits, and protocol-level vulnerabilities further demonstrate that trust and security must be managed across inter-agent messages, tool boundaries, and shared contexts~\cite{AiTM,AgentsUnderSiege,NetSafe,MASLEAK,deng2026secure,south2025position,36S,35S,41S}.

The quality of an agentic service system thus depends not only on task completion but also on collaborative efficiency, state consistency, error propagation depth, recovery capacity, security containment, trust calibration, evidence integrity, and responsibility decomposability. Managing these relationships requires deliberate engineering decisions across the lifecycle: how relationships are described, formed, operated, observed, repaired, and evolved. This establishes relationship-level assurance as a core pillar of ASC.

\section{Agentic Services Computing}\label{sec:asc}

ASC provides a systematic lifecycle framework for engineering agentic services as dependable service entities. Once autonomous behavior becomes a managed service object (Section~\ref{sec:agentic-services}) and participates in goal-driven service systems (Section~\ref{sec:agentic-service-systems}), services computing must address how to represent this behavior before execution, connect it into governable relationships, deliver it under runtime commitments, and preserve accountability through evidence and controlled evolution. ASC extends traditional service engineering, from function-centric endpoints and request-response invocation, to autonomous service behavior that is goal-driven, stateful, tool-mediated, and accountable.

\begin{figure*}[t]
    \centering
    \includegraphics[width=0.8\textwidth]{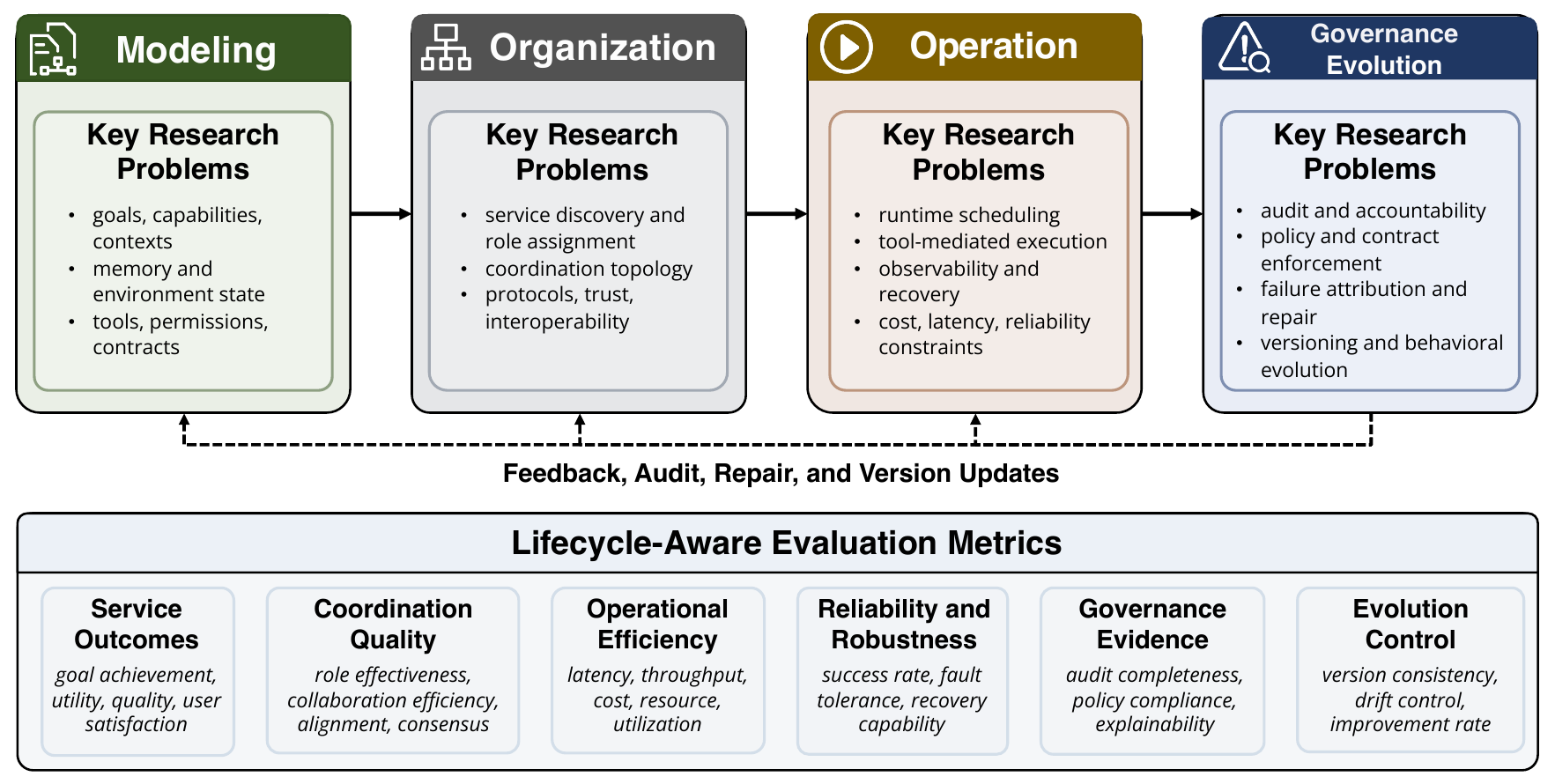}
    \caption{A lifecycle roadmap for Agentic Services Computing. ASC organizes the engineering of agentic services around modeling, organization, operation, and governance and evolution, with lifecycle-aware evaluation metrics spanning service outcomes, coordination quality, operational efficiency, reliability and robustness, governance evidence, and evolution control.}
    \label{fig:asc_roadmap}
\end{figure*}

The ASC lifecycle comprises four interdependent stages. \textit{Modeling} represents agentic services through goals, capabilities, contexts, tools, constraints, contracts, and risk boundaries, making autonomous behavior describable and composable. \textit{Organization} connects multiple agentic services into dynamic service relationships through discovery, role assignment, coordination structures, protocols, trust mechanisms, and responsibility boundaries. \textit{Operation} transforms organized services into runtime service fulfillment, where delegated or authorized goals are interpreted, planned, executed, observed, recovered, and delivered under constraints including latency, cost, reliability, safety, and side effects. \textit{Governance and Evolution} preserves accountability and enables controlled improvement by auditing traces, enforcing policies, attributing failures, managing versions, and regulating behavioral change. Across all stages, \textit{lifecycle-aware evaluation metrics} assess not only QoS or task accuracy but also the service-level quality of autonomous behavior, spanning outcome fulfillment, coordination quality, operational efficiency, robustness, governance evidence, and evolution control. Figure~\ref{fig:asc_roadmap} summarizes this integrated lifecycle roadmap.

\subsection{Modeling}
\label{sec:modeling}

In ASC, modeling defines how agentic services are represented \textit{before} they are discovered, composed, operated, or governed. Traditional service modeling describes interfaces, inputs, outputs, QoS attributes, and deployment states, which are adequate for passive functional endpoints but insufficient for entities that interpret goals, maintain context, invoke tools, coordinate dynamically, and adapt during execution. ASC modeling therefore shifts from interface-level description to service-level representation of autonomous behavior, directly instantiating the service intent, state, and control elements defined in Section~\ref{sec:agentic-services}.

This representation must cover five interconnected dimensions. The \textit{service object} extends functional capability to include delegated goals, contextual assumptions, memory scope, available tools, permissions, quality expectations, behavioral constraints, risk boundaries, and governance interfaces. The \textit{service scenario} specifies the data resources, tool permissions, user roles, organizational boundaries, domain rules, and runtime conditions under which behavior becomes meaningful and valid. The \textit{behavioral chain} captures how a delegated or authorized goal is transformed into task decomposition, tool invocation, interaction, feedback integration, state change, and result delivery. The \textit{contract space} defines allowed actions, prohibited operations, verification requirements, escalation conditions, and human intervention points. Finally, the \textit{autonomy level} indicates when the service may act independently, when supervision is required, and when it should remain advisory. Together, these dimensions make autonomous behavior explicit enough for downstream lifecycle stages.

\subsubsection{Technical Progress}
Recent modeling research has evolved along three converging trajectories that collectively support ASC’s representation needs. First, \textit{capability and context modeling} has expanded beyond operations and QoS to include goals, tools, memory, permissions, and governance interfaces~\cite{zhao2025online,bhardwaj2026agent}. Memory-based, graph-based, and interaction-centered environment models further enrich this by capturing historical context, resource dependencies, and latent interaction factors~\cite{park2025generative,wang2025unveiling,ratnabala2025magnnet,liu2025mosaic,zhang2025sotopia,yue2025masrouter}. In ASC, such models gain service relevance only when anchored to concrete scenarios rather than treated as standalone abstractions.

Second, \textit{temporal behavior modeling} has emerged to make execution processes observable. AgentOps, AgentTrace, and AgentSight demonstrate that goals, plans, tool calls, messages, memory updates, workflow states, and system effects can be represented as traceable service artifacts~\cite{dong2024agentops,alsayyad2026agenttrace,zheng2025agentsight}. These representations provide the evidence foundation for operation and governance, enabling runtime behavior to be checked against service commitments.

Third, \textit{normative modeling} has advanced to specify not just what a service can do, but under what conditions it may act. AgentSLA, Policy Cards, AgentGuard, and AgentProof introduce models of expected outcomes, behavioral constraints, risk levels, tool permissions, verification requirements, and recovery rules~\cite{jouneaux2025agentsla,mavravcic2025policy,koohestani2025agentguard,xavier2026agentproof}. This trajectory extends modeling from descriptive to prescriptive, defining autonomy boundaries and intervention points essential for governable service delivery.

\subsubsection{Modeling Requirements}
These advances point to three foundational requirements for ASC modeling. First, ASC requires a \textit{unified service schema} that integrates goals, capabilities, contexts, tools, permissions, contracts, risks, autonomy levels, and governance interfaces within a single representation. Without such unification, discovery, composition, monitoring, and governance remain fragmented across incompatible descriptions.
Second, modeling must be \textit{scenario-grounded}. The same agentic capability may be useful, unsafe, or irrelevant depending on data resources, tool permissions, user roles, organizational boundaries, and runtime conditions. Service models must therefore bind capability claims to concrete scenarios, linking them to expected value, cost, latency, compliance, and risk exposure.
Third, modeling must support \textit{traceable and contract-aware behavior representation}. Agentic services cannot be modeled solely by intended outcomes; they require explicit representation of behavioral chains, provenance records, constraints, verification points, and autonomy boundaries. This enables subsequent lifecycle stages to execute, observe, recover, audit, and evaluate autonomous behavior under explicit service commitments, directly supporting the operational and governance requirements detailed in Sections~\ref{sec:operation} and~\ref{sec:governance}.

\subsection{Organization}
\label{sec:organization}

In ASC, organization concerns how multiple agentic services are structured, coordinated, and composed to fulfill complex service goals. Traditional service composition is workflow-centric, connecting functional endpoints through predefined invocation orders and data flows. This paradigm is effective when interfaces and execution paths are stable~\cite{9026755}. Agentic services fundamentally alter this problem: because they interpret goals, use tools, maintain context, and adapt during execution, organization must shift from connecting endpoints to forming \textit{dynamic service relationships} among autonomous entities. These relationships encompass roles, capabilities, tasks, contexts, tool dependencies, permissions, trust, cost, and responsibility boundaries.

\begin{figure}[t]
    \centering
    \includegraphics[width=\linewidth]{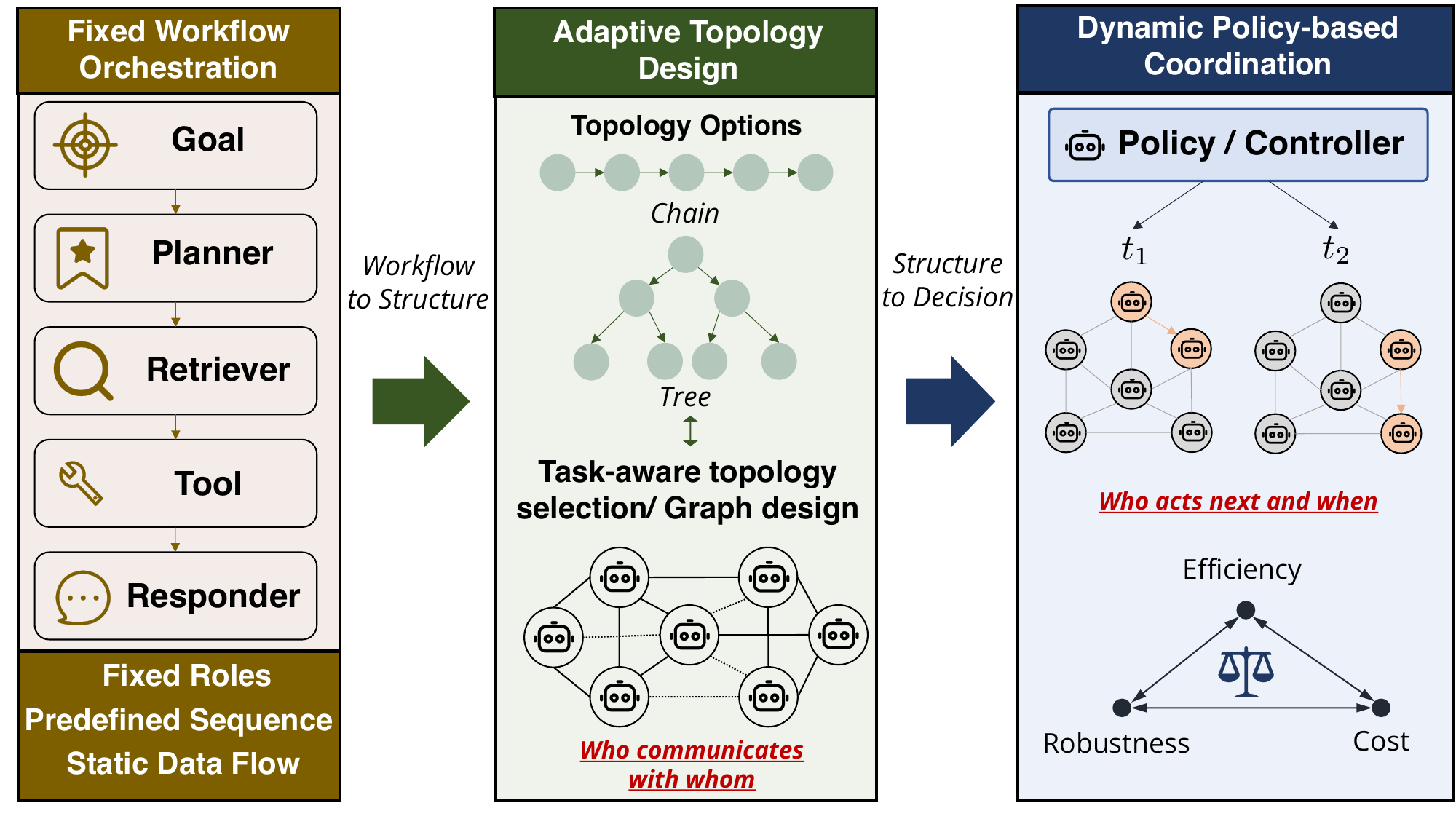}
    \caption{Evolution of ASC organization from fixed workflow orchestration to adaptive topology design and dynamic policy-based coordination.}
    \label{fig:asc-organization-evolution}
\end{figure}

An agentic service system is not a static pipeline but a dynamic service network in which entities interact, negotiate, route tasks, invoke tools, and coordinate under evolving task conditions. Organization must therefore determine how roles are assigned, capabilities matched, tasks routed, communication structured, conflicts arbitrated, and collaboration structures adapted, all under resource constraints, permission boundaries, and risk levels. Figure~\ref{fig:asc-organization-evolution} illustrates this transition from fixed workflow orchestration to adaptive topology design and dynamic policy-based coordination, reflecting the relationship-level assurance principles established in Section~\ref{sec:agentic-service-systems}.

\subsubsection{Technical Progress}
Recent work reveals three layered trajectories in organization research that align with ASC's needs. First, \textit{structural design} has moved beyond fixed orchestration toward task-aware topology construction. While static workflows remain interpretable, they fail when tasks demand changing roles, capabilities, or risk constraints~\cite{li2024survey}. G-Designer and AMAS address this by designing or selecting communication topologies based on task semantics~\cite{GDesigner2025,AMAS2025}, demonstrating that organization quality depends not only on agent availability but on how information and responsibility flow among them.

Second, \textit{runtime coordination control} has emerged to manage interaction dynamics adaptively. Graph-based and robustness-oriented studies show that interaction structure directly affects reproducibility, efficiency, error propagation, and security~\cite{zhang2025unifying,chen2025agentguard,wang2025g}. Policy-driven and sequential orchestration methods like MaAS and AnyMAC treat organization as a runtime decision problem, selecting service configurations or active agents based on task context~\cite{MaAS2025,AnyMAC2025}. These approaches confirm that ASC organization requires both structural design and adaptive control over interaction dynamics.

Third, \textit{open ecosystem organization} extends the scope from internal orchestration to cross-domain service ecosystems. IoA and AgentNet demonstrate registration, discovery, grouping, and autonomous task routing across heterogeneous agents via protocol-based or decentralized mechanisms~\cite{chen2025internet,yang2026agentnet}. This direction introduces additional requirements around trust, accountability, permission exposure, coordination cost, and cross-service responsibility, which directly motivate the relationship-level assurance framework in Section~\ref{sec:agentic-service-systems}. Table~\ref{tab:organization-methods} summarizes representative methods across these trajectories, highlighting their core mechanisms and trade-offs relevant to ASC.

\newcolumntype{L}[1]{>{\raggedright\arraybackslash}p{#1}}
\newlength{\OrgCatW}\newlength{\OrgMethW}\newlength{\OrgCoreW}\newlength{\OrgTradeW}
\setlength{\OrgCatW}{3cm}\setlength{\OrgMethW}{2.1cm}\setlength{\OrgCoreW}{6.5cm}\setlength{\OrgTradeW}{5cm}

\begin{table*}[t]
\centering
\caption{Comparison of representative organization methods for agentic service systems.}
\label{tab:organization-methods}
\scriptsize
\renewcommand{\arraystretch}{0.8}
\setlength{\tabcolsep}{4pt}
\begin{tabular}{@{}L{\OrgCatW} L{\OrgMethW} L{\OrgCoreW} L{\OrgTradeW}@{}}
\toprule
\textbf{Category} & \textbf{Methods} & \textbf{Core Mechanism} & \textbf{Advantages / Limitations} \\
\midrule
Fixed orchestration & \cite{li2024survey} & Static roles, fixed invocation order, predefined flows & Simple and interpretable / Weak adaptability and scalability \\

Topology-aware organization & \cite{GDesigner2025,zhang2025unifying} & Task-aware design of agent communication graphs & Efficient information routing / Sensitive to topology design \\

Dynamic topology selection & \cite{AMAS2025} & Context-dependent selection of collaboration structures & Avoids one-size-fits-all patterns / Extra selection cost and instability risk \\

Automated organization design & \cite{zhang2025agentorchestra,roman2026orchestral} & Automated search over orchestration patterns & Systematic design exploration / Costly search and evaluation dependence \\

Robustness-oriented organization & \cite{chen2025agentguard,wang2025g} & Reliability- and safety-aware interaction analysis & Improved robustness and risk awareness / Increased monitoring overhead \\

Policy-driven configuration & \cite{MaAS2025} & Task-specific sampling of multi-agent configurations & Flexible agent combinations / Search-space dependence and overhead \\

Sequential decision orchestration & \cite{AnyMAC2025} & Context-based selection of next active agent & Adaptive reasoning trajectories / Lower determinism and reproducibility \\

Controller-based evolving orchestration & \cite{EvolvingOrchestrationNeurIPS2025} & Learned or centralized scheduling of interaction flows & Efficient coordination and reduced redundancy / Controller bottleneck risk \\

Open decentralized organization & \cite{chen2025internet,yang2026agentnet} & Registration, discovery, grouping, autonomous routing across heterogeneous agents & Supports open ecosystems / Harder trust, accountability, and control \\
\bottomrule
\end{tabular}
\end{table*}

\subsubsection{Organization Requirements}
These trajectories converge on three essential requirements for ASC organization. First, ASC requires \textit{coordination quality control}: more agents, messages, or interaction rounds do not inherently improve outcomes and may instead introduce redundant reasoning, inconsistent states, and unnecessary cost~\cite{WhyMASFail,RevisitingMAD}. Organization mechanisms must regulate how information is exchanged, validated, aggregated, and terminated.
Second, organization must ensure \textit{relationship safety}. Communication topology and delegation structure directly affect error propagation, attack surfaces, permission exposure, and responsibility boundaries. Highly connected structures may enhance information sharing but amplify unreliable or malicious signals; sparse structures may contain risk but reduce coordination efficiency. ASC must therefore treat topology and delegation as reliability and governance decisions instead of merely performance choices.
Third, organization must enable \textit{adaptive yet accountable coordination}. Dynamic policies improve flexibility, but runtime changes in role assignment, service selection, communication paths, and execution order can undermine predictability and reproducibility. Agentic service systems must keep organization decisions traceable, constrained, and auditable so that adaptive collaboration remains compatible with service commitments.

\subsection{Operation}
\label{sec:operation}

\begin{figure}[t]
    \centering
    \includegraphics[width=\linewidth]{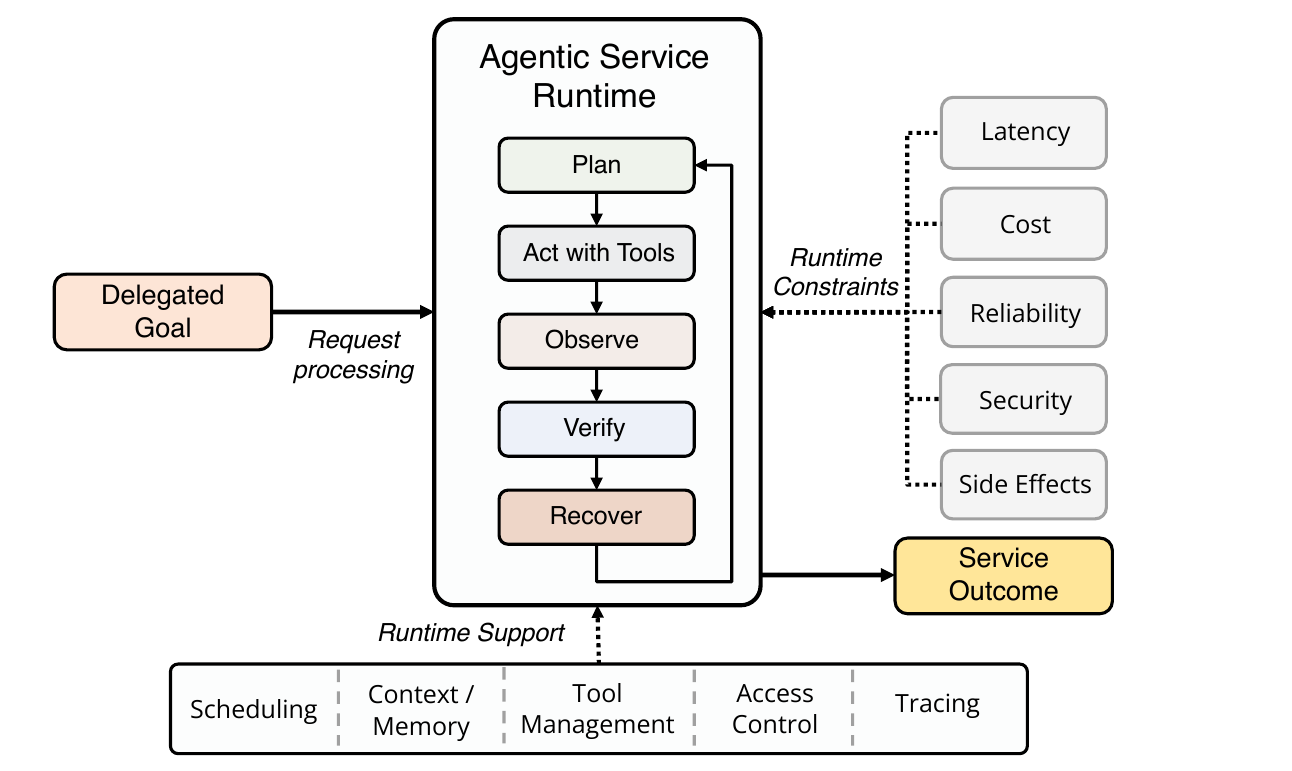}
    \caption{Operational view of agentic services. ASC operation manages the behavioral chain from delegated goals to service outcomes under service commitments related to latency, cost, reliability, security, observability, rollback, and side effects.}
    \label{fig:operation_runtime}
\end{figure}

In ASC, operation concerns how organized agentic services are executed as dependable service workloads and transformed into service outcomes. Traditional service operation primarily manages request-response execution over functional endpoints. It emphasizes availability, concurrency, latency, scaling, recovery, and QoS assurance~\cite{papazoglou2007service,deng2024cloud}. Agentic services preserve these concerns, but the operational object expands from a single invocation to the behavioral chain through which a delegated or authorized goal is fulfilled. This chain includes goal interpretation, planning, tool invocation, state update, inter-agent interaction, exception handling, recovery, and result delivery. Operation in ASC therefore manages goal-driven service fulfillment under runtime commitments. These commitments encompass latency, cost, reliability, security, observability, rollback, and side effects. Figure~\ref{fig:operation_runtime} summarizes this operational view.

\subsubsection{Technical Progress}
Recent progress first appears in runtime substrates that make long-horizon agent execution manageable. AIOS and Agent OS place scheduling, context management, memory management, storage, tool management, and access control into agent operating system abstractions~\cite{mei2025aios,koubaa2025agent}. NALAR treats agent workflows as serving units. It separates workflow specification from execution through managed state and a two-level control plane~\cite{laju2026nalar}. Autellix and ThunderAgent further demonstrate that agent serving must coordinate LLM calls, tool calls, runtime state, KV cache, external resources, and program-level control flow~\cite{2R,kang2026thunderagent}. These systems shift service runtime design from instance-level request scheduling toward behavioral chain management for stateful agentic workflows.

Operational studies in cloud and software environments show how this runtime view supports service fulfillment. Incident diagnosis and mitigation systems such as RCAgent, OpenRCA, TAMO, RCAFlow, STRATUS, FLASH, and K8sGPT connect alerts, telemetry, runbooks, topology information, and operational tools into executable service processes~\cite{wang2024rcagent,xu2025openrca,zhang2025tamo,gao2026rcaflow,chen2025stratus,zhang2024flash,k8sgpt}. Similar patterns appear in code localization, repository auditing, service recommendation, privacy compliance auditing, and industrial on-call automation. In these domains, agents must reason over domain context and act through controlled tools~\cite{chen2025locagent,guo2025repoaudit,liu2026mars,zhang2026priagent,fu2025llm}. These works indicate that operation in ASC is not merely model inference or tool invocation. It is the managed delivery of delegated service work.

Observability and assurance studies make this behavioral chain inspectable during execution. AgentOps identifies goals, plans, prompts, tools, memory, workflows, and feedback as artifacts that should be observed in agent operation~\cite{dong2024agentops}. AgentTrace and AgentSight connect agent-level traces with system-level events. Runtime verification, path policies, exception handling, and trace-based assurance constrain unsafe paths and support recovery during execution~\cite{alsayyad2026agenttrace,zheng2025agentsight,koohestani2025agentguard,kaptein2026runtime,zhou2025shielda,paduraru2026trace}. Consequently, observability becomes an integral part of service execution control rather than a passive record of final outputs.

\subsubsection{Operational Requirements}
These studies reveal three operational requirements for ASC. First, agentic services require a shared runtime abstraction. This abstraction must treat goals, sessions, tools, state, memory, permissions, traces, side effects, and service commitments as connected service objects. Without it, policies and service-level objectives remain tied to prompts, tool calls, or workflows. They become difficult to reuse across platforms.
Second, operation must control fulfillment tradeoffs. Additional reasoning, tools, agents, verification, or human review may improve outcomes. However, they also increase cost, latency, privacy exposure, and failure surfaces. Runtime mechanisms should decide when to reason, act, reuse state, request confirmation, terminate, or recover. These decisions must be based on service criticality, confidence, risk, and cost.
Third, operation must verify state and side effects. Agentic services may update memory, modify code, change cloud configurations, or call external APIs. They therefore require checkpoints, rollback, permission control, isolation, and side-effect validation. Claim-action-state consistency is central to this requirement. Execution hallucination becomes a service-level failure when reported progress, executed actions, and environment records diverge~\cite{xu2025reducing,liu2026agenthallu,zhang2026litmus}. This directly instantiates the relationship-level assurance principles defined in Section~\ref{sec:agentic-service-systems}.

\subsection{Governance and Evolution}
\label{sec:governance}

In ASC, governance and evolution concern how agentic services remain trustworthy, accountable, and improvable after entering service ecosystems. Traditional service maintenance primarily monitors service states, analyzes logs and metrics, detects anomalies, recovers faults, and upgrades versions~\cite{deng2024cloud}. Agentic services extend this scope because their behavior may change through prompts, policies, tools, models, memory, workflows, roles, and collaboration patterns. Their actions may also leave external effects across services or organizations. Governance therefore shifts from maintaining system health to governing autonomous service behavior. It must preserve evidence about goals, decisions, tool use, state changes, side effects, interventions, costs, versions, and responsibility boundaries. Evolution then turns evaluated traces and failures into controlled improvement through versioned, reversible, and policy-aware changes.

\subsubsection{Technical Progress}
Recent work shows a gradual shift from observing final outputs to governing the process through which autonomous service behavior is produced. Table~\ref{tab:governance_evolution_methods} summarizes representative studies. AgentOps and AgentSight illustrate this shift by treating goals, plans, prompts, tool calls, memory updates, workflow states, and system effects as behavioral evidence rather than incidental logs~\cite{dong2024agentops,zheng2025agentsight}. With such evidence, governance can reconstruct how a service goal was interpreted. It can also determine how execution proceeded and which role, tool, message, state update, or intervention shaped the final outcome.

\begin{table*}[t]
    \centering
    \caption{Methods for governing and evolving agentic services.}
    \scriptsize
    \label{tab:governance_evolution_methods}
    \setlength{\belowcaptionskip}{8pt}
    \setlength{\tabcolsep}{4pt}
    \renewcommand{\arraystretch}{1.15}
    \begin{tabular}{@{}p{0.22\textwidth}p{0.74\textwidth}@{}}
        \toprule
        \textbf{Representative Studies} & \textbf{Key Features and Applications} \\
        \midrule
        \cite{dong2024agentops,alsayyad2026agenttrace,zheng2025agentsight}
        & Capture goals, prompts, tool calls, memory updates, workflow steps, and system events; provide behavioral evidence for auditing and accountability. \\
        
        \cite{mavravcic2025policy,jouneaux2025agentsla,bhardwaj2026agent}
        & Encode policies, quality expectations, behavioral constraints, and recovery rules; turn governance requirements into service-level artifacts. \\
        
        \cite{xavier2026agentproof,koohestani2025agentguard,kaptein2026runtime,yuan2026aegis}
        & Verify or enforce tool use, execution paths, and safety constraints before or during execution; reduce unsafe autonomous behavior. \\
        
        \cite{WhyMASFail,xue2025characterization,zhu2025llm,ma2025diagnosing,shah2026characterizing}
        & Identify failures from roles, memory, planning, tools, workflows, framework bugs, and external service interactions; motivate artifact-level governance. \\
        
        \cite{30R,31R,barke2026agentrx,zhou2025shielda,47R}
        & Localize failures, attribute responsibility, debug trajectories, handle exceptions, and support repair of multi-agent service workflows. \\
        
        \cite{li2025agentgit,liu2025towards}
        & Extend versioning to prompts, tools, workflows, memory, models, and execution environments; support rollback and controlled updates. \\
        
        \cite{wu2025evolver,zhang2026memskill,huang2026ama,zhu2026hybrid}
        & Distill experience, evolve skills and memory, and improve behavior over time; require evaluation and governance to avoid behavioral drift. \\
        
        \cite{yang2025agentic,guo2026skillprobe,sharma2025argent}
        & Expose governance issues across service ecosystems, including identity, provenance, third-party tools, skill marketplaces, and cross-domain accountability. \\
        \bottomrule
    \end{tabular}
\end{table*}

This evidence becomes useful only when connected to executable boundaries. Policy Cards and Agent Behavioral Contracts suggest that governance artifacts are moving from external documentation toward runtime constraints. These constraints cover goals, tools, actions, risk levels, recovery rules, and accountability evidence~\cite{mavravcic2025policy,bhardwaj2026agent}. Related work on AgentProof, AgentGuard, runtime path policies, and trace-based assurance demonstrates the same tendency from another angle~\cite{xavier2026agentproof,koohestani2025agentguard,kaptein2026runtime,paduraru2026trace}. Governance is increasingly inserted into the execution path. Permissions, obligations, and unsafe trajectories can thus be checked before service side effects become difficult to reverse.

The same process view also changes how failures are handled. Studies on multi-agent failure and agent hallucination show that errors may arise from role design, memory, planning, action selection, tool use, workflow orchestration, framework behavior, external services, or their interactions~\cite{WhyMASFail,xue2025characterization,zhu2025llm,ma2025diagnosing,liu2026agenthallu}. Attribution and debugging methods therefore move the unit of analysis from the final answer to the trajectory~\cite{28R,ge2025introducing,29R,west2025abduct,barke2026agentrx,wang2026flat}. They help locate the artifact or relation that causes a deviation. This capability is especially important when execution hallucination appears as a mismatch among claims, actions, and environment states~\cite{zhang2026litmus}.

Governance also supports service evolution. AgentGit and versioning studies show that change management for agentic services cannot stop at code or deployment artifacts. Prompts, tools, workflows, memory, models, data, and execution environments may all alter service behavior~\cite{li2025agentgit,liu2025towards}. Experience-based improvement and memory evolution further demonstrate how execution traces can be reused to improve skills and behavior~\cite{wu2025evolver,zhang2026memskill,huang2026ama,zhu2026hybrid}. Evolution is useful but also risky. Improvements must remain evaluated, comparable, reversible, and consistent with policies. In open ecosystems, the same concern extends to identity, provenance, trust, reputation, and cross-organizational accountability when agentic services use third-party tools or shared skills~\cite{yang2025agentic,fang2025we,guo2026skillprobe,58R,sharma2025argent}.

\subsubsection{Governance Requirements}
These advances point to three requirements for ASC. Governance requires continuous behavioral evidence across the service lifecycle. Goals, plans, messages, tool calls, state transitions, memory updates, side effects, human interventions, and versions should be recorded in forms that support audit, attribution, recovery, and evolution.
Governance also requires runtime boundaries. Policies, contracts, permissions, risk levels, cost limits, and human confirmation rules should be connected to execution control rather than checked only after completion. This connection allows autonomous behavior to be constrained before unsafe actions or irreversible side effects occur. It also provides direct feedback to the operation stage for real-time fulfillment tradeoff control.
Evolution requires controlled change. Agentic services may improve through prompt updates, model changes, tool changes, memory updates, workflow revision, and feedback learning. These changes should be versioned, evaluated, regression tested, and reversible. Service behavior can thus improve without losing its accountable identity.

\subsection{Lifecycle-Aware Evaluation Metrics}
\label{sec:metrics}

The ASC lifecycle fundamentally changes how service quality is evaluated. Traditional service systems are commonly assessed through functional correctness, availability, latency, throughput, reliability, cost, and QoS or SLA satisfaction~\cite{zeng2003quality,jatoth2015computational,ghafouri2020survey,deng2024cloud}. Agent benchmarks often emphasize task accuracy or success rate. These measures remain useful, but they do not fully capture whether autonomous behavior can be delivered as a dependable service. ASC therefore extends evaluation toward service-level behavior fulfillment quality. The evaluation object is the behavioral chain and service relationship network through which a delegated goal is interpreted, executed, governed, recovered, and improved.

This view can be organized around five metric families. Outcome metrics evaluate whether the delegated goal is fulfilled and whether the result creates service value for a domain task such as diagnosis, mitigation, recommendation, auditing, or compliance~\cite{chen2025stratus,zhang2025tamo,liu2026mars,guo2025repoaudit,zhang2026priagent}. Coordination metrics evaluate whether roles, communication, topology, trust, and aggregation support effective collaboration among service entities~\cite{MultiAgentBench,WhyMASFail,RevisitingMAD,GDesigner2025,AMAS2025,MaAS2025,AnyMAC2025}. Operational metrics evaluate whether fulfillment satisfies runtime commitments such as latency, cost, resource use, tool calls, scheduling overhead, and human intervention~\cite{jouneaux2025agentsla,mei2025aios,laju2026nalar,2R,kang2026thunderagent}.

Reliability and robustness metrics evaluate whether intermediate failures, unsafe actions, side effects, and environmental changes can be detected, contained, recovered, and regression tested. For agentic services, this family must include consistency-oriented measures. Examples include claim-action consistency, action-state verification coverage, reported progress mismatch, invalid tool parameters, unverified side effects, and recovery after consistency violations~\cite{xu2025reducing,zhang2026litmus,chen2025stratus,zhou2025shielda,47R}. These metrics distinguish service-level hallucination from ordinary answer-level hallucination. They check whether autonomous behavior remains grounded in authorized goals and observable environment states.

Governance and evolution metrics evaluate whether service behavior remains visible, compliant, attributable, and improvable. They include trace completeness, audit coverage, policy compliance, permission safety, accountability accuracy, intervention effectiveness, behavior drift, version reproducibility, rollback success, and compatibility with dependent services~\cite{dong2024agentops,alsayyad2026agenttrace,zheng2025agentsight,mavravcic2025policy,bhardwaj2026agent,koohestani2025agentguard,xavier2026agentproof,paduraru2026trace,li2025agentgit,liu2025towards,wu2025evolver,zhang2026memskill,huang2026ama}. Evolution is desirable only when improvements are traceable, comparable, reversible, and consistent with service constraints.

Together, these metrics extend service evaluation from QoS and task accuracy to the service-level quality of autonomous behavior. A high-quality agentic service must not only complete tasks but also coordinate effectively, satisfy runtime commitments, recover from failures, preserve governance evidence, and evolve under controlled change. With this metric view, the ASC lifecycle forms a complete service engineering loop. Modeling makes autonomous behavior representable. Organization connects service entities into service relationships. Operation turns delegated goals into runtime fulfillment. Governance and evolution preserve accountability and controlled improvement. Metrics judge whether the loop produces dependable service value. The next issue is the middleware substrate that can support this loop across heterogeneous protocols, tools, runtimes, and service ecosystems.

\section{Middleware and Enabling Infrastructure for Agentic Services}
\label{sec:middleware}

The ASC lifecycle requires middleware that makes agentic services accessible, executable, and manageable in real service environments. Traditional service middleware supports interface exposure, invocation, message exchange, discovery, traffic control, and deployment management. These functions remain necessary but insufficient. Agentic services additionally involve goals, capabilities, tools, context, execution state, permissions, and behavioral evidence. Middleware in ASC therefore provides the interface and runtime foundation through which autonomous service entities interact with service ecosystems and operate under operational control.

\begin{figure*}[t]
    \centering
    \includegraphics[width=0.8\textwidth]{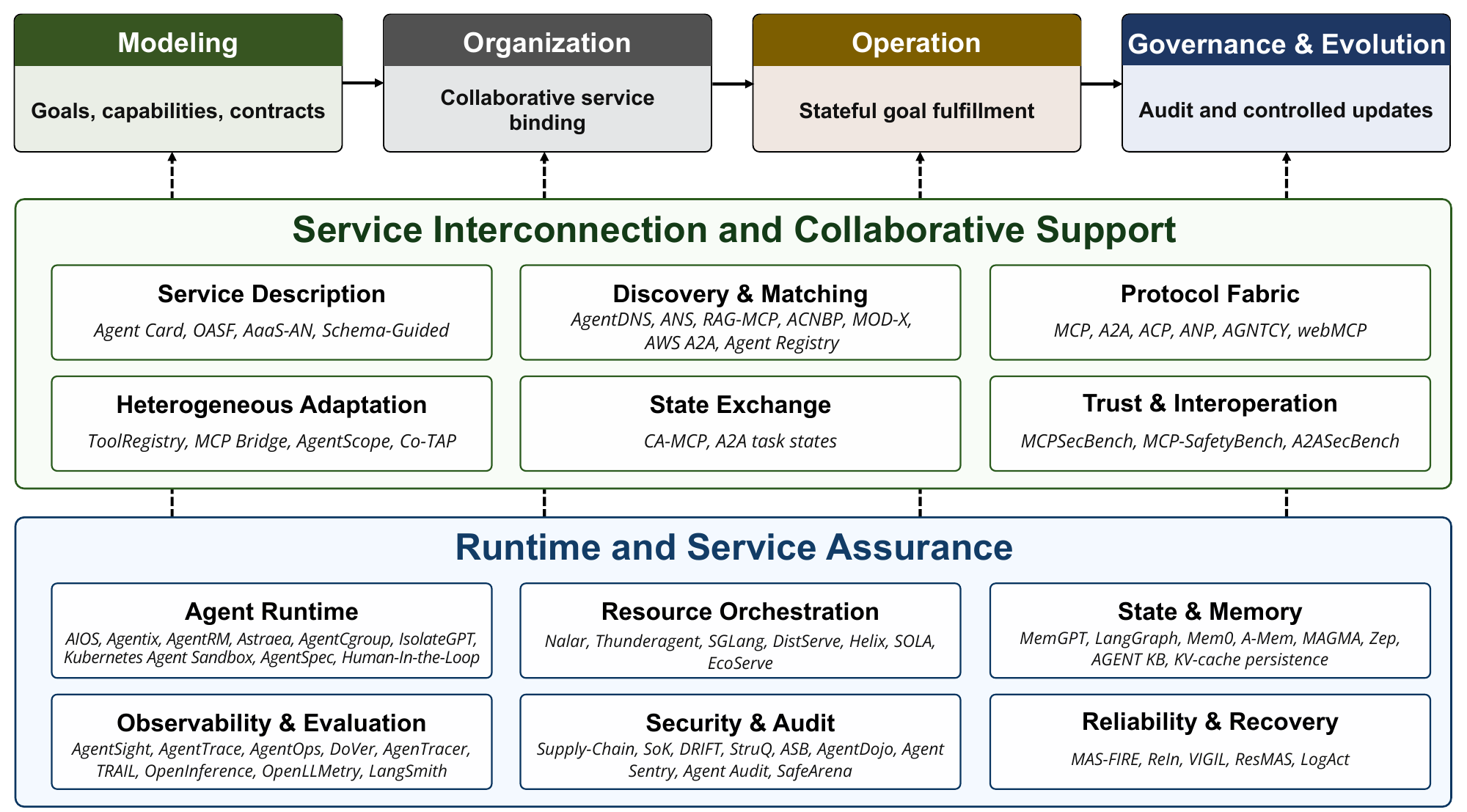}
    \caption{Middleware and enabling infrastructure for agentic services. The interconnection plane provides interface and protocol support for description, discovery, tool access, agent collaboration, adaptation, state exchange, and trusted interaction. The runtime plane provides system support for agent execution, serving, scheduling, isolation, persistence, observability, security, and recovery.}
    \label{fig:middleware}
\end{figure*}

Figure~\ref{fig:middleware} summarizes this infrastructure through two support planes. The service interconnection and collaborative support plane provides interface and protocol support for description, discovery, tool access, agent collaboration, adaptation, state exchange, and trusted interaction. The runtime and service assurance plane provides system support for agent execution, serving, scheduling, isolation, persistence, observability, security, and recovery. Together, these planes instantiate the modeling, organization, operation, and governance requirements defined in Section~\ref{sec:asc}.

\subsection{Service Interconnection and Collaborative Support}
\label{sec:interconnection}

The interconnection layer defines how agentic services expose their capabilities and interact with tools, agents, and external service environments. Conventional service interfaces mainly describe operations and data schemas. Agentic service interfaces must additionally represent goals, skills, tool access, authentication requirements, action boundaries, and failure expectations. Recent interface description studies reflect this shift from operation-oriented contracts toward capability- and behavior-oriented contracts~\cite{a2a,2S,zhu2025agent,10S}. Such descriptions make autonomous service behavior understandable before invocation.

Discovery and protocol support form the next part of this interface layer. Service requests may be expressed as delegated goals. Discovery must therefore match intent with agent capabilities, tool availability, permission conditions, and trust metadata~\cite{4S,5S,9S,6S,gao2025task}. Emerging protocols further separate several interaction planes. Tool access protocols expose resources and executable functions to agents. Agent collaboration protocols support task-oriented communication and state tracking. Federation and directory protocols support cross-domain discovery and connection~\cite{mcp,a2a,18S,17S,16S,2S,13S,14S}. These efforts suggest a layered interface architecture for ASC. Tool invocation, agent communication, and federation operate under consistent assumptions about identity, state, permission, and audit evidence.

Interconnection middleware must also connect agentic services with existing software assets and trusted interaction contexts. Many useful capabilities remain exposed through REST APIs, OpenAPI specifications, enterprise systems, local tools, and framework-specific connectors. Work on protocol adaptation, API wrapping, specification generation, and unified agent interfaces shows how these assets can be exposed to agents through controlled access~\cite{23S,24S,29S,30S,28bS,25S,26S}. Task states, context identifiers, shared context, scoped authorization, consent mechanisms, and protocol security studies demonstrate additional requirements. Agentic service interfaces must support session continuity, access control, and audit-ready interaction records~\cite{a2a,32S,34S,35S,36S,37S,38S,41S,42S}. The interconnection plane therefore provides the interface foundation for describing, discovering, connecting, and trusting agentic services in heterogeneous service ecosystems.

\subsection{Runtime and Service Assurance}
\label{sec:runtime}

The runtime and service assurance layer provides the system substrate for deploying and operating agentic services as manageable workloads. Agent operating systems and serving engines form the basic execution foundation. They place scheduling, context management, memory management, tool management, storage, access control, resource allocation, and state-aware execution into runtime abstractions for agents~\cite{mei2025aios,2R,4R,5R,3R,koubaa2025agent}. Agent serving systems further treat multi-step agent programs as execution units. They coordinate model inference with tool calls, runtime state, external resources, and workflow control~\cite{laju2026nalar,kang2026thunderagent,zheng2024sglang,zhong2024distserve,mei2025helix,hong2025sola}. These systems indicate that ASC needs portable runtime primitives such as sessions, tools, state, memory, resources, and permissions.
F
Engineering frameworks, isolation mechanisms, and persistence layers provide practical support for building service-grade agentic systems. Existing frameworks support persistent multi-actor workflows, multi-agent communication, role-based collaboration, tool-connected execution, and human review points~\cite{LangGraph,12R,wu2024autogen,hong2024metagpt,wang2025openhands}. Managed sessions, sandboxed execution, policy-based privileges, and runtime specifications constrain agent execution when agents edit files, run code, call APIs, query databases, or operate cloud resources~\cite{6R,7R,9R,10R,11R}. Resource orchestration is becoming a key runtime concern for agentic services, requiring workflow-aware, program-aware, heterogeneous, and sustainability-aware scheduling beyond request-level model serving~\cite{laju2026nalar,kang2026thunderagent,zheng2024sglang,zhong2024distserve,mei2025helix,hong2025sola,li2025ecoserve,moore2025sustainable}. Persistence and memory infrastructure preserve workflow progress, retain useful context, recover sessions, and manage long-horizon state under operational control~\cite{14R,13R,12R,15R,16R,17R,18R,19R,locomo2024,longmemeval2025,memgym2026,contextEngineering2025,agenticContextEngineering2026}.

Observability, debugging, and reliability tools complete the runtime support plane. Agent-specific tracing records prompts, messages, tool calls, memory updates, workflow steps, and system effects. Debugging and telemetry tools connect these traces with diagnosis and monitoring ecosystems~\cite{zheng2025agentsight,alsayyad2026agenttrace,dong2024agentops,34R,35R,36R,37R}. Reliability mechanisms based on fault injection, runtime supervision, intervention, recovery, fallback, retry, and routing support controlled failure handling~\cite{47R,49R,50R,51R,portkey,litellm}. The current runtime landscape remains fragmented across operating systems, serving engines, frameworks, sandboxes, memory layers, and observability tools. A key middleware challenge for ASC is to standardize portable primitives such as session, tool, state, memory, permission, trace, checkpoint, sandbox, and recovery policy. This standardization enables agent prototypes to evolve into service-grade agentic systems and directly supports the application domains discussed next.

\section{Applications of Agentic Services}
\label{sec:applications}

Agentic services become practically meaningful when autonomous behavior is embedded in operational settings. These settings provide service goals, controlled tool access, runtime state, service entry points, measurable outcomes, and governance evidence. Agents in these environments do not merely respond to user prompts. They receive delegated or authorized goals, interact with domain tools, maintain execution context, produce verifiable outcomes, and leave behavioral traces. These traces support monitoring, debugging, auditing, and improvement. Application domains therefore serve as critical evidence for understanding how autonomous behavior transforms into service delivery.

\subsection{Serviceization Dimensions}
\label{sec:app_dimensions}

A clear view of application maturity emerges when the focus shifts from task completion to service delivery. Early demonstrations often evaluate agents by their ability to solve bounded tasks. Service settings require connecting the same tasks with domain data, tool permissions, execution state, verification evidence, and governance records. Across current applications, this serviceization appears through six recurring dimensions. A service goal may be delegated by a user or triggered by an authorized event. Tools and environments provide the action space through which the agent creates value. Runtime state records the continuity of task context, memory, intermediate results, and recovery points. Service entries connect the agent with workflows, platforms, APIs, repositories, or operational signals. Outcomes make the service value measurable and verifiable. Governance evidence supports audit, attribution, and controlled improvement. These dimensions connect application evidence with the ASC lifecycle. They clarify why application maturity depends on more than task accuracy alone.

\subsection{Domain Evidence for Service Delivery Maturity}
\label{sec:app_domains}

Cloud operations, software engineering, enterprise workflow, recommendation, and Web fulfillment demonstrate varying degrees of service delivery maturity. In AIOps and IT service management, systems such as OpenRCA, TAMO, RCAFlow, STRATUS, AIOpsLab, ITBench, and K8sGPT connect incident goals with logs, metrics, traces, runbooks, topology information, and operational tools~\cite{xu2025openrca,zhang2025tamo,gao2026rcaflow,chen2025stratus,ma2025aiopslab,jha2025itbench,k8sgpt}. These systems approach service delivery because they originate from operational events and depend on telemetry, tool use, execution trajectories, and evaluable outcomes.

Software engineering illustrates the transition from model capability to delegated service process most clearly. Code generation was initially exposed as a local snippet completion feature. Recent agentic coding systems instead organize software work as a service. A user issue, feature request, bug report, or audit objective becomes a service goal. The repository, dependency graph, tests, shell, development environment, and pull request workflow constitute the service environment. The agent navigates files, modifies code, executes commands, observes test feedback, revises patches, and produces reviewable changes. Research systems like SWE-agent, AutoCodeRover, OpenHands, LocAgent, and RepoAudit demonstrate this pattern~\cite{yang2024sweagent,Zhang2024,wang2025openhands,chen2025locagent,guo2025repoaudit}. Industrial platforms including GitHub Copilot coding agent, OpenAI Codex, Claude Code, OpenClaw, and Harness AI show its adoption in practical engineering environments~\cite{github2025copilotcodingagent,openai2025introducingcodex,anthropic2026claudecodeoverview,openclaw2026homepage,harness2026agents}. The service outcome includes a trajectory from intent to patch, together with repository state, tool access records, test evidence, and correction paths. Agentic coding thus represents a stateful and verifiable service process.

Enterprise and Web-facing settings embed agentic services in business and digital environments. WorkArena and $\tau$-bench place agents in tasks that depend on tools, users, policies, and workflow state~\cite{AlexandreDrouin2024,shunyu2025bench,boisvert2024workarena++}. Microsoft Copilot Studio, Dynamics 365, ServiceNow, Agentforce, Oracle Fusion Agentic Applications, and UiPath connect agents with business objects, event triggers, approval paths, and audit-oriented controls~\cite{microsoft2025copilotStudioEventTriggers,microsoft2026dynamicsSalesAgents,servicenow2025yokohamaAIAgents,salesforceAgentforceGuide,salesforceAgentforceTrustLayer,oracle2026fusionAgenticApplications,uipath2025AgenticAutomationPlatform}. In recommendation and Web fulfillment, MARS, MACRec, WebArena, and VisualWebArena show agents refining user intent, inspecting digital states, selecting actions, and completing service tasks through interfaces~\cite{liu2026mars,wang2024macrec,zhou2024webarena,koh2024visualwebarena}. Successful delivery in these settings depends on process constraints, user context, explicit service entry points, and verifiable completion.

\begin{figure}[t]
    \centering
    \includegraphics[width=1.0\columnwidth]{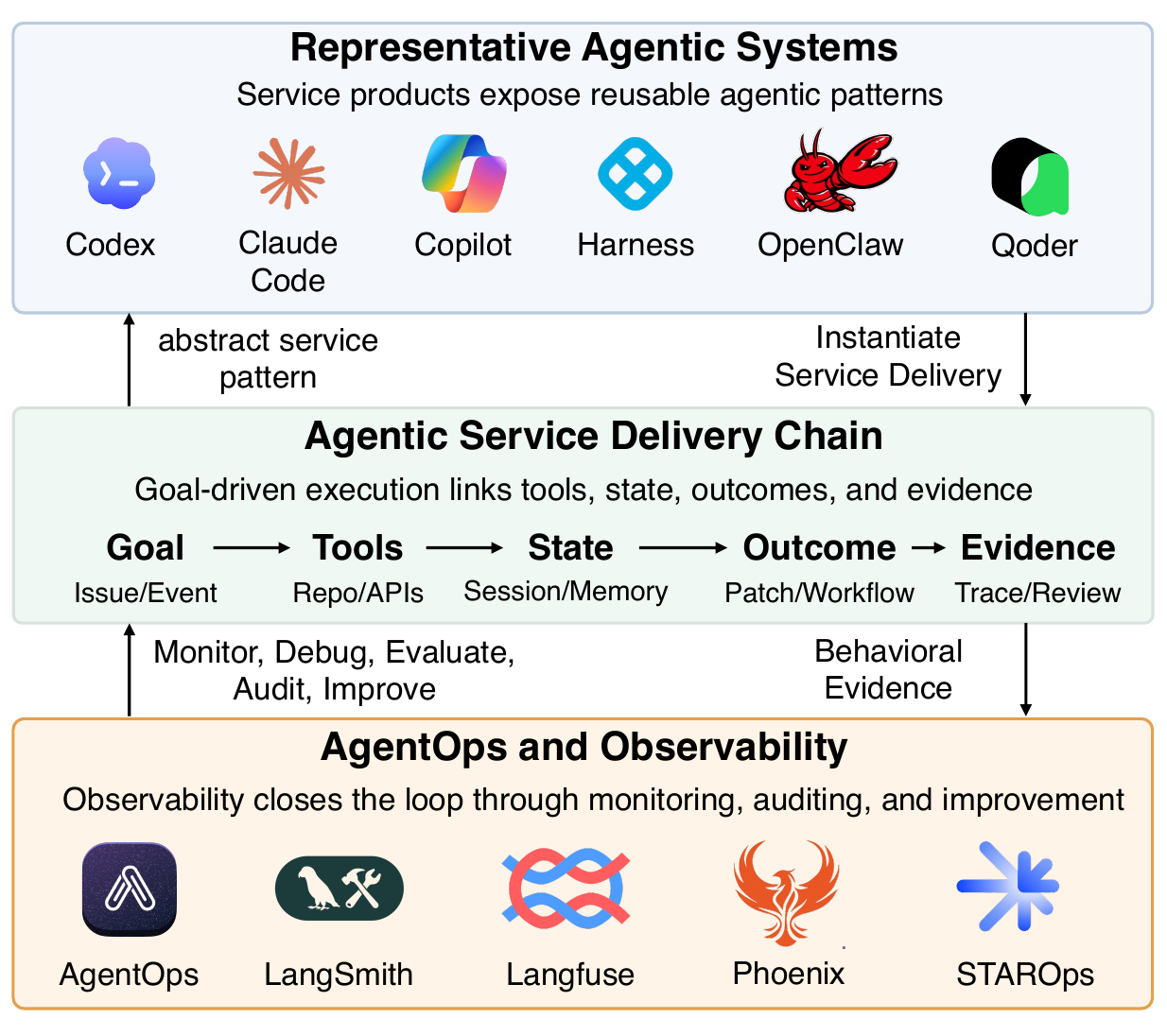}
    \caption{Representative industrial evidence for agentic serviceization. Agentic systems instantiate service delivery chains. Observability platforms consume behavioral evidence and provide feedback for monitoring, debugging, evaluation, audit, and improvement.}
    \label{fig:application_serviceization}
\end{figure}

\begin{table*}[t]
    \centering
    \caption{Serviceization evidence in representative application domains of ASC. \ding{51} denotes strong evidence; $\Delta$ denotes partial or emerging evidence.}
    \label{tab:application_serviceization}
    \setlength{\belowcaptionskip}{8pt}
    \scriptsize
    \setlength{\tabcolsep}{2.8pt}
    \renewcommand{\arraystretch}{1.12}
    \begin{tabular*}{0.95\textwidth}{@{\extracolsep{\fill}}p{0.14\textwidth}p{0.43\textwidth}cccccc@{}}
        \toprule
        \textbf{Domain} & \textbf{Representative Evidence} &
        \makecell{\textbf{Goal}} &
        \makecell{\textbf{Tools}} &
        \makecell{\textbf{State}} &
        \makecell{\textbf{Entry/} \\ \textbf{Trigger}} &
        \makecell{\textbf{Outcome/} \\ \textbf{Value}} &
        \makecell{\textbf{Gov.}} \\
        \midrule
        Cloud / AIOps / ITSM
        & OpenRCA~\cite{xu2025openrca}, TAMO~\cite{zhang2025tamo}, RCAFlow~\cite{gao2026rcaflow}, STRATUS~\cite{chen2025stratus}, AIOpsLab~\cite{ma2025aiopslab}, ITBench~\cite{jha2025itbench}, K8sGPT~\cite{k8sgpt}
        & \ding{51} & \ding{51} & \ding{51} & $\Delta$ & \ding{51} & $\Delta$ \\
        Software Eng. / DevOps
        & SWE-agent~\cite{yang2024sweagent}, AutoCodeRover~\cite{Zhang2024}, OpenHands~\cite{wang2025openhands}, LocAgent~\cite{chen2025locagent}, RepoAudit~\cite{guo2025repoaudit}, Copilot~\cite{github2025copilotcodingagent}, Codex~\cite{openai2025introducingcodex}, Claude Code~\cite{anthropic2026claudecodeoverview}, OpenClaw~\cite{openclaw2026homepage}, Harness AI~\cite{harness2026agents}
        & \ding{51} & \ding{51} & \ding{51} & \ding{51} & \ding{51} & $\Delta$ \\
        Enterprise / Workflow
        & WorkArena~\cite{AlexandreDrouin2024}, WorkArena++~\cite{boisvert2024workarena++}, $\tau$-bench~\cite{shunyu2025bench}, Copilot/Dynamics~\cite{microsoft2025copilotStudioEventTriggers,microsoft2026dynamicsSalesAgents}, ServiceNow~\cite{servicenow2025yokohamaAIAgents}, Agentforce~\cite{salesforceAgentforceGuide,salesforceAgentforceTrustLayer}, Oracle Fusion~\cite{oracle2026fusionAgenticApplications}, UiPath~\cite{uipath2025AgenticAutomationPlatform}
        & \ding{51} & \ding{51} & \ding{51} & \ding{51} & \ding{51} & $\Delta$ \\
        Recommendation / Web
        & MARS~\cite{liu2026mars}, MACRec~\cite{wang2024macrec}, WebArena~\cite{zhou2024webarena}, VisualWebArena~\cite{koh2024visualwebarena}
        & \ding{51} & \ding{51} & \ding{51} & $\Delta$ & \ding{51} & $\Delta$ \\
        Privacy / Compliance
        & PriAgent~\cite{zhang2026priagent}, AudAgent~\cite{zheng2026audagentautomatedauditingprivacy}, SkillProbe~\cite{guo2026skillprobe}, EUROCOMPLY~\cite{EURECOM+8507}, Agent Audit~\cite{58R}
        & \ding{51} & \ding{51} & $\Delta$ & $\Delta$ & \ding{51} & \ding{51} \\
        High-Risk / Embodied
        & FinRobot~\cite{yang2024finrobotopensourceaiagent}, Hippocratic AI~\cite{hippocraticai2026homepage}, Coscientist~\cite{Boiko2023}, ChemCrow~\cite{MBran2024}, AI Scientist~\cite{Lu2026}, AutoRT~\cite{ahn2024autortembodiedfoundationmodels}, RoboCasa365~\cite{nasiriany2026robocasa365largescalesimulationframework}
        & \ding{51} & \ding{51} & $\Delta$ & $\Delta$ & $\Delta$ & $\Delta$ \\
        \bottomrule
    \end{tabular*}
\end{table*}

Privacy, compliance, high-risk, and embodied domains extend serviceization to constrained settings. PriAgent, AudAgent, SkillProbe, EUROCOMPLY, and Agent Audit inspect policies, applications, skills, or runtime evidence to produce audit or compliance artifacts~\cite{zhang2026priagent,zheng2026audagentautomatedauditingprivacy,guo2026skillprobe,EURECOM+8507,58R}. FinRobot, Hippocratic AI, Coscientist, ChemCrow, AI Scientist, AutoRT, and RoboCasa365 connect agents with financial analysis, healthcare interaction, scientific instruments, or physical environments~\cite{yang2024finrobotopensourceaiagent,hippocraticai2026homepage,Boiko2023,MBran2024,Lu2026,ahn2024autortembodiedfoundationmodels,nasiriany2026robocasa365largescalesimulationframework}. These domains are less mature as deployable service ecosystems. They nevertheless expose decisive requirements for grounding, oversight, safety boundaries, liability, validation, and evidence preservation when autonomous behavior reaches sensitive or regulated contexts.

\subsection{Observability as an Application-Level Service Capability}
\label{sec:app_observability}

Long execution trajectories create a second application need. Planning, tool use, memory updates, interaction, and recovery themselves become part of service delivery. AgentOps, AgentTrace, and AgentSight treat goals, plans, prompts, messages, tool calls, memory updates, workflow states, and system effects as behavioral evidence~\cite{dong2024agentops,alsayyad2026agenttrace,zheng2025agentsight}. This evidence allows developers and providers to reconstruct goal interpretation, tool invocation sequences, state changes, failure locations, and outcome consistency. Observability thus transcends implementation support. It becomes an application-level service capability that makes autonomous behavior monitorable, debuggable, auditable, and improvable. Such systems complement domain applications by providing the operational evidence needed to manage them as services.

\subsection{Synthesis of Serviceization Evidence}
\label{sec:app_synthesis}

Table~\ref{tab:application_serviceization} summarizes representative domains across the six serviceization dimensions. Figure~\ref{fig:application_serviceization} presents industrial agentic systems as concrete instances of service delivery chains. Observability platforms consume behavioral evidence and provide feedback for continuous improvement. Across these domains, agentic services become distinct when delegated goals connect with controlled tool access, persistent state, explicit entries, measurable outcomes, and governance records. Future studies should advance from bounded task demonstrations to scenario-level service delivery. This requires incorporating domain data, workflow integration, operational commitments, cost bounds, escalation rules, side-effect control, audit records, and feedback loops. These elements will make autonomous behavior comparable, selectable, operable, and improvable as a service offering.

\section{Future Directions}
\label{sec:future}

ASC opens several research directions for services computing by treating autonomous behavior as a service object. The shift from functional endpoints to agentic services requires new ways to describe, contract, operate, govern, and evaluate service entities. Their behavior unfolds through goals, tools, memory, state, interaction, and adaptation. Four research directions align with the ASC lifecycle.

\paragraph{Service abstractions and contracts for autonomous behavior}
Future agentic services need descriptions that connect delegated goals, capabilities, context, memory, tool access, permissions, behavioral constraints, and governance interfaces. These descriptions should support discovery, composition, verification, monitoring, and evolution across the service lifecycle. A related need is service contracts and service-level agreements tailored for agentic services. Such contracts must cover outcome commitments, cost and latency expectations, tool permissions, safety boundaries, escalation rules, rollback conditions, and responsibility allocation.

\paragraph{Lifecycle-native middleware and runtime support}
Agentic services require infrastructures that manage sessions, tools, memory, state, traces, checkpoints, permissions, and side effects as service-level objects. Future middleware should integrate interconnection, collaborative execution, isolation, observability, recovery, and policy enforcement. Autonomous service behavior can then be delivered under operational commitments across platforms and environments.

\paragraph{Behavioral governance and accountability}
Agentic services may inspect private data, modify software artifacts, invoke external services, or coordinate with other agents. Future service ecosystems therefore need mechanisms for identity, provenance, trust, permission delegation, policy-aware tool use, audit, intervention, and rollback. These mechanisms should make intermediate goals, messages, tool calls, state transitions, memory updates, and human interventions accountable throughout execution.

\paragraph{Service-level evaluation beyond traditional QoS}
Availability, latency, and cost remain important. They must however be combined with measures of goal fulfillment, coordination quality, recovery, trace completeness, policy compliance, responsibility attribution, behavioral drift, and continuous improvement. This agenda ensures that evaluation captures the full service delivery quality rather than isolated task accuracy.

\section{Conclusion}
\label{sec:conclusion}

LLM-based agents are changing the nature of service entities. Traditional services computing remains the foundation for description, discovery, composition, delivery, monitoring, and governance. Its function-centered paradigm nevertheless offers limited support for service behavior that is goal-driven, tool-using, stateful, adaptive, and accountable. Agentic Services Computing addresses this shift by placing autonomous behavior within a service-centered framework.
This article defined agentic services and agentic service systems. It analyzed their core elements and service-oriented properties. It further organized related research through lifecycle management, metrics, middleware, applications, and future directions. Through this view, ASC extends services computing from managing reusable functional capabilities to engineering autonomous service entities. These entities can be described, composed, operated, governed, evaluated, and evolved in future digital service ecosystems.

\bibliographystyle{IEEEtran}
\bibliography{ref}


 





\end{document}